\documentstyle[preprint,aps,eqsecnum]{revtex}
\tighten
\begin{document}
\draft
\title{\bf  FRACTAL DIMENSIONS AND SCALING LAWS IN THE INTERSTELLAR
MEDIUM AND GALAXY DISTRIBUTIONS:  A NEW FIELD THEORY APPROACH}
\author{{\bf  H. J. de Vega$^{(a)}$,
  N. S\'anchez$^{(b)}$ and  F. Combes$^{(b)}$,}\bigskip}

\bigskip

\address
{ (a)  Laboratoire de Physique Th\'eorique et Hautes Energies,
Universit\'e Paris VI, Tour 16, 1er \'etage, 4, Place Jussieu
75252 Paris, Cedex 05, FRANCE. Laboratoire Associ\'e au CNRS UA 280.\\
(b) Observatoire de Paris,  Demirm, 61, Avenue de l'Observatoire,
75014 Paris,  FRANCE. 
Laboratoire Associ\'e au CNRS UA 336, Observatoire de Paris et
\'Ecole Normale Sup\'erieure.  \\ }

\date{November 1997}
\maketitle
\begin{abstract}
We develop a field theoretical approach to the cold interstellar medium (ISM)
and large structure of the universe. We show that a non-relativistic
self-gravitating gas in thermal equilibrium with variable
number of atoms or fragments is exactly equivalent to a field theory 
of a single scalar field $ \phi({\vec x}) $ with exponential self-interaction.
We analyze this field theory perturbatively and non-perturbatively
through the renormalization group approach. We show {\bf scaling}  
behaviour (critical) for a continuous range of the temperature and of 
the other 
physical parameters. We derive in this framework the scaling relation 
$ M(R) \sim R^{d_H} $ for the mass on a region of size $ R $,
and $ \Delta v \sim R^q $ for the velocity dispersion where 
$ q = \frac12(d_H -1) $. For the density-density correlations
we find a power-law  behaviour for large distances 
$ \sim |{\vec r_1} -{\vec r_2}|^{2 d_H -6} $.  The fractal dimension
$  d_H $ turns to be related with the critical exponent $ \nu $ of the 
correlation length by $  d_H = 1/ \nu $. The renormalization group
approach for a single component scalar field in three dimensions
states that the long-distance critical behaviour may be governed by the
(non-perturbative) Ising fixed point. The Ising 
values of the scaling exponents are  $  \nu = 0.631...  , \; d_H = 1.585... $
and $ q = 0.293...$. Mean field theory yields for the scaling
exponents $ \nu = 1/2  , \; d_H = 2 $ and $ q = 1/2 $. Both the Ising
and the mean field values are compatible with the present ISM
observational data: $  1.4    \leq   d_H    \leq   2     ,   \;
0.3  \leq     q  \leq 0.6 \;  $.
As typical in critical phenomena, the scaling behaviour and critical
exponents of the ISM can be obtained without dwelling into the dynamical
(time-dependent) behaviour. 
We develop a field theoretical approach to the galaxy distribution. We
consider a gas of self-gravitating masses on the FRW background,
 in quasi-thermal equilibrium. We show that it exhibits
scaling behaviour by renormalization group methods. The galaxy
correlations are first computed  assuming homogeneity for very large
scales and then without assuming homogeneity. In the first case we
find $  \xi(r) \equiv <\rho({\vec r_0})\rho({\vec r_0} + {\vec r}) > /
<\rho>^2-1 \sim r^{-\gamma} $, with $ \gamma = 2 $. In the second
case we find  $ D(r) = <\rho({\vec r_0})\rho({\vec r_0} + {\vec r}) >
\; \sim\;  r^{-\Gamma} $ with $  \Gamma = 1 $.
While the universe becomes more and more homogeneous at large scales,
statistical analysis of galaxy catalogs have revealed a fractal structure 
at small-scales ($\lambda < 100 h^{-1}$ Mpc), with a fractal dimension
$D=1.5-2$ (Sylos Labini et al 1996). 

We study the thermodynamics of a
self-gravitating system with the theory of critical phenomena and finite-size
scaling and show that gravity provides a dynamical mechanism to produce 
this fractal structure. Only a
limited, (although large),  range of scales is involved, between a 
short-distance cut-off
below which other physics intervene, and a large-distance cut-off,
where the thermodynamic equilibrium is not satisfied. The galaxy ensemble can
be considered at critical conditions, with large density fluctuations
developping at any scale. From the theory of critical phenomena, 
we derive the two independent critical exponents $\nu$ and $\eta$ and 
predict the fractal dimension $ D = 1/\nu $  to be either $1.585$ or $2$,
depending on whether the long-range behaviour is governed by 
the Ising or the mean field fixed points, respectively. Both set of values
are compatible with present observations. In addition,
we predict the scaling behaviour of the gravitational
potential to be  $ r^{- \frac12 (1 + \eta )} $. That is, $ r^{-0.5} $
for mean field or $ r^{- 0.519} $ for the Ising fixed point. The theory allows
to compute the three and higher density correlators without any assumption
or Ansatz. We find that the connected $N$-points density scales as 
$ r_1^{N(D-3)} $, when $ r_1 >> r_i, \; 2\leq i \leq N $. There are no free 
parameters in this theory.
\end{abstract}
\section{The Self-Interacting Gravitational Gas}

We review recent work \cite{natu,prd,gal} on the statistical properties of
a self-interacting gravitational gas in thermal equilibrium. We
discuss two relevant astrophysical applications of the
self-interacting gravitational gas: the cold interstellar medium (ISM) and
the galaxy distributions.

This review is organized as follows. In section II we summarize the 
main properties of  the ISM and our results for it, in section III 
we develop the field
theory approach to the gravitational gas. A short distance cutoff is
naturally present here and prevents zero distance gravitational
collapse singularities (which would be unphysical in the present
case). Here, the cutoff theory is physically meaningful. The
gravitational gas is also treated in a $D$-dimensional space.
In section IV we study the scaling behaviour and thermal fluctuations
both in perturbation theory and non-perturbatively (renormalization
group approach). $g^2 \equiv \mu\, T_{eff} $ acts as the dimensionless
coupling constant for the non-linear fluctuations of the field $\phi$.
We show that the non-linear fluctuations of the field $\phi$ 
make the theory to scale ( critical behaviour) for a continuous range of 
values of the parameters of the theory. Thus, changing  the parameters 
keeps the theory at {\bf criticality}. The renormalization group analysis 
made in section IV confirm such results. External forces to
the ISM gas like stars are shown {\bf not} to affect the scaling behaviour
of the gas. That is, the scaling exponents $ q , \; d_H $ are solely
governed by fixed points and hence, they are stable under perturbations.

In section V we discuss the galaxy distributions starting from the
its observational status, in section VI we  clarify the definition of the 
galaxy correlators and then in section VII we
 develop the dynamical equations in the 
comoving frame. We  apply our field-theory approach in section VIII. We
derive in this framework two and three point correlators deriving the 
scaling exponents. Discussion and remarks are presented in section IX. 


\section{The Interstellar Medium: Introduction and results}

The interstellar medium (ISM) is a gas essentially formed by atomic (HI) 
and molecular ($H_2$) hydrogen, distributed in cold ($T \sim 5-50 K$) 
clouds, in a very inhomogeneous and fragmented structure. 
These clouds are confined in the galactic plane 
and in particular along the spiral arms. They are distributed in 
a hierarchy of structures, of observed masses from 
$1\; M_{\odot}$ to $10^6 M_{\odot}$. The morphology and
kinematics of these structures are traced by radio astronomical 
observations of the HI hyperfine line at the wavelength of 21cm, and of
the rotational lines of the CO molecule (the fundamental line being
at 2.6mm in wavelength), and many other less abundant molecules.
  Structures have been measured directly in emission from
0.01pc to 100pc, and there is some evidence in VLBI (very long based 
interferometry) HI absorption of structures as low as $10^{-4}\; pc = 20$ AU 
(3 $10^{14}\; cm$). The mean density of structures is roughly inversely
proportional to their sizes, and vary 
between $10$ and $10^{5} \; atoms/cm^3$ (significantly above the 
mean density of the ISM which is about 
$0.1 \; atoms/cm^3$ or $1.6 \; 10^{-25}\; g/cm^3$ ).
Observations of the ISM revealed remarkable relations between the mass, 
the radius and velocity dispersion of the various regions, as first 
noticed by Larson \cite{larson}, and  since then confirmed by many other 
independent observations (see for example ref.\cite{obser}). 
From a compilation of well established samples of data for many different  
types of molecular clouds of maximum linear dimension (size) $ R $,  
total mass $M$ and internal velocity dispersion $ \Delta v$ in each region: 
\begin{equation}\label{vobser}
M (R)  \sim    R^{d_H}     \quad        ,     \quad  \Delta v \sim R^q \; ,
\end{equation}
over a large range of cloud sizes, with   $ 10^{-4}\; - \; 10^{-2}
\;  pc \;   \leq     R   \leq 100\;  pc, \;$
\begin{equation}\label{expos}
1.4    \leq   d_H    \leq   2     ,   \;     0.3  \leq     q  \leq
0.6 \; . 
\end{equation}
These {\bf scaling}  relations indicate a hierarchical structure for the 
molecular clouds which is independent of the scale over the above 
cited range; above $100$ pc in size, corresponding to giant molecular clouds,
larger structures will be destroyed by galactic shear.

These relations appear to be {\bf universal}, the exponents 
$d_H , \; q$ are almost constant over all scales of the Galaxy, and
whatever be  
the observed molecule or element. These properties of interstellar cold 
gas are supported first at all from observations (and for many different 
tracers of cloud structures: dark globules using $^{13}$CO, since the
more abundant isotopic species $^{12}$CO is highly optically thick, 
dark cloud cores using $HCN$ or $CS$ as density tracers,
 giant molecular clouds using $^{12}$CO, HI to trace more diffuse gas, 
and even cold dust emission in the far-infrared).
Nearby molecular clouds are observed to be fragmented and 
self-similar in projection over a range of scales and densities of 
at least $10^4$, and perhaps up to $10^6$.

The physical origin as well as the interpretation of the scaling relations 
 (\ref{vobser}) are not theoretically understood. 
The theoretical derivation of these
 relations has been the subject of many proposals and controversial 
discussions. It is not our aim here to account for all the proposed models 
of the ISM and we refer the reader to refs.\cite{obser} for a review.

The physics of the ISM is complex, especially when we consider the violent
perturbations brought by star formation. Energy is then poured into 
the ISM either mechanically through supernovae explosions, stellar winds,
bipolar gas flows, etc.. or radiatively through star light, heating or
ionising the medium, directly or through heated dust. Relative velocities
between the various fragments of the ISM exceed their internal thermal
speeds, shock fronts develop and are highly dissipative; radiative cooling
is very efficient, so that globally the ISM might be considered 
isothermal on large-scales. 
Whatever the diversity of the processes, the universality of the
scaling relations suggests a common mechanism underlying the physics.

  We propose that self-gravity is the main force at the origin of the 
structures, that can be perturbed locally by heating sources. 
Observations are compatible with virialised structures at all scales.
 Moreover, it has been suggested that the molecular clouds ensemble is
in isothermal equilibrium with the cosmic background radiation at $T \sim 3 K$
in the outer parts of galaxies, devoid of any star and heating
sources \cite{pcm}. This colder isothermal medium might represent the ideal
frame to understand the role of self-gravity in shaping the hierarchical
structures. Our aim is to show that the scaling laws obtained are then
quite stable to perturbations.

Till now, no theoretical derivation of the scaling laws 
eq.(\ref{vobser}) has been 
provided in which the values of the exponents are {\bf obtained} from the 
theory (and not just taken from outside or as a starting input or hypothesis).

The aim of our work \cite{natu,prd,gal} 
 is to develop a theory of the cold ISM. A first 
step in this goal is to provide a theoretical derivation of the scaling 
laws eq.(\ref{vobser}), in which the values of the exponents $d_H , \; q$ are
{\bf obtained} from the theory.
For this purpose, we will implement for the ISM the powerful tool of field 
theory and the Wilson's approach to critical phenomena \cite{kgw}.
 
We  consider a gas of non-relativistic atoms interacting with each other
 through Newtonian gravity and which are in thermal
equilibrium at temperature $ T $.
We  work in the grand canonical ensemble, allowing for a variable
number of particles $N$.

Then, we show that this system 
is exactly equivalent to a field theory of a single scalar field
$\phi({\vec x})$ with 
exponential  interaction. We express the grand canonical partition function
$ {\cal Z} $ as
\begin{equation}\label{zetafi}
{\cal Z} =  \int\int\;  {\cal D}\phi\;  e^{-S[\phi(.)]} \; , 
\end{equation}
where
\begin{eqnarray}\label{SmuyT}
S[\phi(.)] & \equiv &  {1\over{T_{eff}}}\;
\int d^3x \left[ \frac12(\nabla\phi)^2 \; - \mu^2 \; 
e^{\phi({\vec x})}\right] \; , \cr \cr
 T_{eff} &=& 4\pi \; {{G\; m^2}\over {T}} \quad , \quad
\mu^2 = \sqrt{2\over {\pi}}\; z\; G \, m^{7/2} \, \sqrt{T} \; ,
\end{eqnarray}
$ m $ stands for the mass of the atoms and $ z $  for the fugacity. 
We show that in the $\phi$-field language, the particle density 
expresses as
\begin{equation}\label{denfi}
  <\rho({\vec r})> =  -{1 \over {T_{eff}}}\;<\nabla^2 \phi({\vec r})>=
{{\mu^2}\over{T_{eff}}} \; <e^{\phi({\vec r})}>  \; .
\end{equation}
where $ <\ldots > $ means functional average over   $ \phi(.) $
with statistical weight $  e^{-S[\phi(.)]} $. Density correlators are
written as
\begin{eqnarray}\label{correI}
C({\vec r_1},{\vec r_2})&\equiv&
<\rho({\vec r_1})\rho({\vec r_2}) > -<\rho({\vec r_1})><\rho({\vec r_2}) > 
\cr \cr
&=&  {{\mu^4}\over{T_{eff}}^2} \; \left[  
<e^{\phi({\vec r_1})} \; e^{\phi({\vec r_2})}> -
<e^{\phi({\vec r_1})}> \; <e^{\phi({\vec r_2})}> \right]\; .
\end{eqnarray}
The  $\phi$-field defined by eqs.(\ref{zetafi})-(\ref{SmuyT}) has remarkable
properties under  scale transformations 
$$
{\vec x} \to {\vec x}_{\lambda} \equiv \lambda{\vec x} \; ,
$$
where $\lambda$ is an arbitrary real number. For any solution  $
\phi({\vec x}) $ of the stationary point equations,
\begin{equation}\label{eqMovI}
\nabla^2\phi({\vec x}) +  \mu^2 \; e^{\phi({\vec x})} = 0 \; ,
\end{equation}
there is a family of dilated solutions of the same equation (\ref{eqMovI}),
given by
$$
\phi_{\lambda}({\vec x}) \equiv \phi(\lambda{\vec x}) +\log\lambda^2
\; .
$$
In addition, $ S[\phi_{\lambda}(.)] = \lambda^{2-D} \; S[\phi(.)] $. 

\bigskip

We study the field theory (\ref{zetafi})-(\ref{SmuyT}) both
perturbatively and non-perturbatively.

The computation of the thermal fluctuations through the evaluation of
the functional integral  eq.(\ref{zetafi}) is quite non-trivial. We
use the scaling property as a guiding principle. In order to built a
perturbation theory in the dimensionless coupling  $ g \equiv
\sqrt{\mu \, T_{eff}} $ we look for stationary points of
eq.(\ref{SmuyT}). We compute the density correlator eq.(\ref{correI}) to
leading order in $ g $. For large distances it behaves as

\begin{equation}\label{coRa}
 C({\vec r_1},{\vec r_2}) \buildrel{  | {\vec  r_1} - {\vec  r_2}|\to
\infty}\over =  {{ \mu^4 }\over {32\, \pi^2 \; 
 | {\vec  r_1} - {\vec  r_2}|^2}} + O\left( \; | {\vec
r_1} - {\vec r_2}|^{-3}\right)\; . 
\end{equation}

We  analyze further this theory with the renormalization group
approach. Such non-perturbative approach is the more powerful
framework to
derive scaling behaviours in field theory \cite{kgw,dg,nn}.

We show that the mass contained in a region of size $ R $ scales as
$$
<M(R)> =    m   \; \int^R
<e^{\phi({\vec x})}> \; d^3x   \simeq  m  \, V \, a +m  \, {K
\over{1-\alpha}}\; R^{ \frac1{\nu}} + \ldots\; , 
$$
and the mass fluctuation, $ (\Delta M(R))^2 = <M^2>-<M>^2 $,
scales as
$$
\Delta M(R)  \sim  R^{d_H}\; .
$$
Here $ \nu $ is the correlation length critical exponent for the
$\phi$-theory (\ref{zetafi}) and $ a $ and $ K $ are  constants. Moreover, 
\begin{equation}\label{SdensiI}
<\rho({\vec r})> = m  a \; + m \, {K
\over{4\pi\nu(1-\alpha)}}\;r^{ \frac1{\nu}-3} \quad {\rm for}\; r
\;  {\rm of ~ order}\;\sim R \; .
\end{equation}

The scaling exponent $\nu$ can be identified with the inverse
Haussdorf (fractal) dimension $d_H$ of the system
$$
d_H = \frac1{\nu} \; .
$$
In this way, $ M \sim R^{d_H} $ according to the usual  definition of fractal
dimensions \cite{sta}.

In Fourier space,
\begin{equation}\label{fou}
 <{\tilde \rho}({\vec k})> \sim k^{ -\frac1{\nu}} \; .
\end{equation}


From the renormalization group analysis, 
the density-density correlators (\ref{correI}) result to be, 
\begin{equation}\label{corI}
C({\vec r_1},{\vec r_2})\sim |{\vec r_1} -{\vec r_2}|^{\frac2{\nu} -6} \; .
\end{equation}
Computing the average gravitational potential energy and using the
virial theorem yields for the velocity dispersion,
$$
\Delta v \sim R^{\frac12(\frac1{\nu} -1)} \; .
$$
This gives a new scaling relation between the exponents $ d_H $ and $ q $
$$
q =\frac12\left(\frac1{\nu} -1\right) =\frac12(d_H -1)  \; .
$$

The perturbative calculation (\ref{coRa}) yields the mean field value
for $ \nu $ \cite{ll}. That is,
\begin{equation}\label{meanF}
 \nu= \frac12  \quad ,  \quad d_H = 2 \quad {\rm and } \quad q = \frac12 \; .
\end{equation}
 
We find scaling behaviour in the $\phi$-theory for a {\bf continuum set} of
values of $\mu^2$ and $ T_{eff} $.
The renormalization group transformation amounts to replace 
the parameters $ \mu^2 $ and $ T_{eff} $ 
in $ \beta\, H $ and $ S[\phi(.)] $ by the effective ones at the scale
$ L $ in question. 

The renormalization group approach applied to a
{\bf single}  component scalar field in three space dimensions
indicates that the long distance critical behaviour is governed by the
(non-perturbative) Ising fixed point  \cite{kgw,dg,nn}. 
Very probably, there are no further fixed points \cite{grexa}. 
The scaling exponents associated to the Ising fixed point are

\begin{equation}\label{Isint}
\nu = 0.631...  \quad , \quad d_H = 1.585...   \quad {\rm and} \quad
 q = 0.293...\; \; .
\end{equation}

Both the mean field  (\ref{meanF}) and the Ising  (\ref{Isint}) numerical 
values are compatible with the present observational values  
(\ref{vobser}) - (\ref{expos}).

\bigskip

The theory presented here also predicts a power-law behaviour for
the two-points ISM density correlation functionl (see eq.(\ref{corI}),
$ 2 d_H - 6 = - 2.830\ldots$, for the Ising fixed point 
and $ 2 d_H - 6 = - 2 $ for the mean field exponents),
that should be compared with observations. Previous attempts to
derive correlation functions from observations were not entirely conclusive, 
because of lack of dynamical range \cite{klein}, but much more extended maps of
the ISM could be available soon to test our theory. In addition, we predict 
an independent exponent for the gravitational 
potential correlations ($ \sim r^{-1-\eta} $, where
$ \eta_{Ising}=0.037\ldots $ and $ \eta_{mean ~ field} = 0 $
\cite{dg}), which could be checked through 
gravitational lenses observations in front of quasars.

\bigskip

The mass parameter $\mu $ [see eq.(\ref{SmuyT})] in the $\phi$-theory
turns to coincide at the tree level with the inverse of the Jeans length
\begin{equation}\label{lonJeans}
\mu =  \sqrt{12 \over {\pi}}\; { 1 \over {d_J}} \; .
\end{equation}

We find that in the scaling domain  the Jeans distance $ d_J $ grows
as $  <d_J> \sim R $. This shows that the Jeans distance   {\bf scales} 
with the  {\bf size} of the system and therefore the instability is
 present for all sizes $ R $. Had  $ d_J $ being  of order larger than
$ R $, the Jeans instability would be absent. 

\bigskip

The gravitational gas in thermal equilibrium explains quantitatively
the observed scaling laws in the ISM. This fact does not exclude  
turbulent phenomena in the ISM. 
Fluid flows (including turbulent regimes) are probably relevant
in the dynamics (time dependent processes) of the ISM. As usual in critical
phenomena \cite{kgw,dg}, the equilibrium scaling laws  can be  understood 
for the ISM without  dwelling with the dynamics. 
A further step in the study of the ISM will be to include the
dynamical (time dependent) description within the field theory
approach presented in this review.

\bigskip

If the ISM is considered as a flow,  the Reynolds number $Re_{ISM}$ on
scales  $L \sim 100$pc  has a very high value of the order of $10^6$.  
This led to the suggestion that the ISM (and the universe in general)
could be {\bf modelled}  as a turbulent flow \cite{weisz}. 
(Larson \cite{larson} first observed that the 
exponent in the power-law relation for the velocity dispersion is not greatly 
different from the Kolmogorov value $1/3$ for subsonic turbulence).

It must be noticed that the turbulence hypothesis for the ISM is based on 
the comparison of the ISM with the results known for incompressible 
flows. However, the physical conditions in the ISM are 
very different from those of incompressible flows in the laboratory. 
(And the 
study of ISM turbulence needs more complete and enlarged investigation 
than those performed until now based in the concepts of flow turbulence 
in the laboratory).  
Besides the facts that the ISM exhibits large density fluctuations on all 
scales, and the observed fluctuations are highly supersonic, (thus the 
ISM can not viewed as an `incompressible' and  `subsonic' flow),
 and besides other differences, an essential feature to point out is that
 the long-range self-gravity interaction present in the ISM is completely 
absent in the studies of flow turbulence. 
In any case, in a satisfactory theory of the ISM, 
it should be possible to extract the  behaviours of 
the ISM (be turbulent or whatever) from the theory 
as a result, instead to be introduced as a starting input or  hypothesis.

\section{Field theory approach to the gravitational gas}

Let us  consider a gas of non-relativistic atoms with mass $m$ interacting
only through Newtonian gravity and which are in thermal
equilibrium at temperature $ T \equiv \beta^{-1} $.
We shall work in the grand canonical ensemble, allowing for a variable
number of particles $N$.

The grand partition function of the system can be written as

\begin{equation}\label{gfp}
{\cal Z} = \sum_{N=0}^{\infty}\; {{z^N}\over{N!}}\; \int\ldots \int
\prod_{l=1}^N\;{{d^3p_l\, d^3q_l}\over{(2\pi)^3}}\; e^{- \beta H_N}
\end{equation}
where
\begin{equation}\label{hami3}
H_N = \sum_{l=1}^N\;{{p_l^2}\over{2m}} - G \, m^2 \sum_{1\leq l < j\leq N}
{1 \over { |{\vec q}_l - {\vec q}_j|}}
\end{equation}
$G$ is Newton's constant and $z$ is the fugacity.

The integrals over the momenta $p_l, \; (1 \leq l \leq N) $
can be performed explicitly in eq.(\ref{gfp}) 
using
$$
\int\;{{d^3p}\over{(2\pi)^3}}\; e^{- {{\beta p^2}\over{2m}}} =
\left({m \over{2\pi \beta}}\right)^{3/2}
$$
We thus find,
\begin{equation}\label{gfp2}
\displaystyle{
{\cal Z} = \sum_{N=0}^{\infty}\; {1 \over{N!}}\;
\left [ z\left({m \over{2\pi \beta}}\right)^{3/2}\right]^N
\; \int\ldots \int
\prod_{l=1}^N d^3q_l\;\; e^{ \beta G \, m^2 \sum_{1\leq l < j\leq N}
{1 \over { |{\vec q}_l - {\vec q}_j|}} }}
\end{equation}

We proceed now to recast this many-body problem into a field theoretical
form \cite{origen,stra,sam,kh}.
 
Let us define the density
\begin{equation}\label{defro}
\rho({\vec r})= \sum_{j=1}^N\; \delta({\vec r}- {\vec q}_j)\; ,
\end{equation}
such that, we can rewrite the potential energy in eq.(\ref{gfp2}) as
\begin{equation}\label{PotE}
 \frac12 \, \beta G \, m^2 \sum_{1\leq l \neq j\leq N}
{1 \over { |{\vec q}_l - {\vec q}_j|}} =  \frac12\,  \beta \, G \, m^2
\int_{ | {\vec x} - {\vec y}|> a}\;
{{d^3x\, d^3y}\over { | {\vec x} - {\vec y}|}}\; \rho({\vec x})
\rho({\vec y}) \; .
\end{equation}
The cutoff $ a $ in the r.h.s. is introduced in order to avoid
self-interacting divergent terms. However, such divergent terms would
contribute to ${\cal Z}$ by
an infinite multiplicative factor that can be factored out.

By using
$$ 
\nabla^2 { 1 \over { | {\vec x} - {\vec y}|}}= -4\pi \; \delta( {\vec
x} - {\vec y}) \; ,
$$
and partial integration we can now represent the exponent of the
potential energy eq.(\ref{PotE}) as a functional integral\cite{stra}
\begin{equation}\label{reprf}
e^{  \frac12\, \beta G \, m^2
\int \;
{{d^3x\, d^3y}\over { | {\vec x} - {\vec y}|}}\; \rho({\vec x})
\rho({\vec y})} = \int\int\; {\cal D}\xi \; e^{ -\frac12\int d^3x \; (\nabla
\xi)^2 \; + \; 2 m \sqrt{\pi G\beta}\; \int d^3x \; \xi({\vec x})\;
\rho({\vec x}) } 
\end{equation}

Inserting this expression into eq.(\ref{gfp2})  and using
eq.(\ref{defro}) yields 
\begin{eqnarray}\label{gfp3} 
{\cal Z} &=& \sum_{N=0}^{\infty}\; {1 \over{N!}}\;
\left [ z\left({m \over{2\pi \beta}}\right)^{3/2}\right]^N\;
 \int\int\;  {\cal D}\xi \; e^{ -\frac12\int d^3x \; (\nabla \xi)^2}
\; \int\ldots \int
\prod_{l=1}^N d^3q_l\; \; e^{ 2 m  \sqrt{\pi G\beta}\; \sum_{l=1}^N
\xi({\vec q}_l)} \cr \cr
 &=& \int\int\;  {\cal D}\xi \; e^{ -\frac12\int d^3x  \;(\nabla \xi)^2}\;
 \sum_{N=0}^{\infty}\; {1 \over{N!}}\;
\left [ z\left({m \over{2\pi \beta}}\right)^{3/2}\right]^N\;
\left[ \int d^3q \;  e^{ 2 m  \sqrt{\pi G\beta}\;\xi({\vec q})}
\right]^N \cr \cr
 &=& \int\int\;  {\cal D}\xi \; e^{ -\int d^3x \left[ \frac12(\nabla \xi)^2\;
- z \left({m \over{2\pi \beta}}\right)^{3/2}\; e^{ 2 m \sqrt{\pi
G\beta}\;\xi({\vec x})}\right]} \; \quad .
\end{eqnarray}

It is convenient to introduce the dimensionless field
\begin{equation}
\phi({\vec x}) \equiv  2 m \sqrt{\pi G\beta}\;\xi({\vec x}) \; .
\end{equation}

Then,
\begin{equation}\label{zfi}
{\cal Z} =  \int\int\;  {\cal D}\phi\;  e^{ -{1\over{T_{eff}}}\;
\int d^3x \left[ \frac12(\nabla\phi)^2 \; - \mu^2 \; e^{\phi({\vec
x})}\right]}\; , 
\end{equation}

where
\begin{equation}\label{muyT}
\mu^2 = \sqrt{2\over {\pi}}\; z\; G \, m^{7/2} \, \sqrt{T} 
\quad , \quad T_{eff} = 4\pi \; {{G\; m^2}\over {T}} \; .
\end{equation}
The partition function for the gas of particles in gravitational
interaction has been transformed into the partition function for a single
scalar field $\phi({\vec x})$  with  {\bf local} action
\begin{equation}\label{acci}
S[\phi(.)] \equiv  {1\over{T_{eff}}}\;
\int d^3x \left[ \frac12(\nabla\phi)^2 \; - \mu^2 \; e^{\phi({\vec
x})}\right] \; .
\end{equation}
The  $\phi$  field exhibits an exponential self-interaction $ - \mu^2
\; e^{\phi({\vec x})} $. 

Notice that the effective 
temperature $ T_{eff} $ for the  $\phi$-field partition function 
turns out to be  {\bf inversely}
proportional to $ T $ whereas the characteristic length $\mu^{-1}$ behaves
as $ \sim T ^{-1/4}$. This is a duality-type mapping between the two models.

It must be noticed that the term $ - \mu^2 \; e^{\phi({\vec x})} $
makes the  $\phi$-field energy density unbounded from
below. Actually, the initial Hamiltonian (\ref{gfp}) is also  unbounded from
below. This unboundness physically originates in the attractive
character of the gravitational force. Including a short-distance
cutoff [see sec. 2A, below] eliminates the zero distance singularity
and hence the possibility  of zero-distance
collapse which is unphysical in the present context. 
We therefore expect meaningful physical results in the
cutoff theory. Moreover, assuming zero boundary conditions for
$\phi({\vec r})$ at $ r \to \infty $ shows that the derivatives of
$\phi$ must also be large if $ e^\phi$ is large. Hence, the term $
\frac12(\nabla\phi)^2 $ may stabilize the energy.

The action  (\ref{acci}) defines a non-renormalizable field
theory for any number of dimensions $ D > 2 $.
This is a further reason to keep the short-distance cutoff non-zero.

\bigskip

Let us compute now the statistical average  value of the density
$\rho({\vec r})$ which in  the grand canonical ensemble is given by
\begin{equation}
<\rho({\vec r})> =   {\cal Z}^{-1}\; \sum_{N=0}^{\infty}\; {1 \over{N!}}\;
\left [ z\left({m \over{2\pi \beta}}\right)^{3/2}\right]^N
\; \int\ldots \int
\prod_{l=1}^N d^3q_l\; \; \rho({\vec r}) \; 
e^{ \frac12\, \beta G \, m^2 \sum_{1\leq l \neq j\leq N}
{1 \over { |{\vec q}_l - {\vec q}_j|}} }\; .
\end{equation}

As usual in the functional integral calculations, 
it is convenient to introduce sources in the partition function (\ref{zfi})
in order to compute  average values of fields

\begin{equation}\label{zfiJ}
{\cal Z}[J(.)] \equiv  \int\int\; {\cal D}\phi\;  e^{ -{1\over{T_{eff}}}\;
\int d^3x \left[ \frac12(\nabla
\phi)^2 \; - \mu^2 \; e^{\phi({\vec x})}\; \right]
+\int d^3x  \;J({\vec x})\; \phi({\vec x}) \; }\; .
\end{equation}
The average value of $ \phi({\vec r}) $ then writes as
\begin{equation}
< \phi({\vec r})> = {{\delta \log{\cal Z} }\over{\delta J({\vec r})}}\; .
\end{equation}

In order to compute $<\rho({\vec r})>$ it is useful to introduce
\begin{equation}
{\cal V}[J(.)] \equiv  \frac12 \,\beta G \, m^2
\int_{ | {\vec x} - {\vec y}|> a }\;
{{d^3x\, d^3y}\over { | {\vec x} - {\vec y}|}}\; 
\left[ \rho({\vec x})+ \;J({\vec x})\;\right]
\left[\rho({\vec y})+ \;J({\vec y})\;\right]\; .
\end{equation}
Then, we have
$$
 \rho({\vec r}) \; e^{{\cal V}[0]} = -{1 \over{T_{eff}}} \; \nabla^2_{\vec
x} \left({{\delta}\over{\delta J({\vec r})}} e^{{\cal V}[J(.)]}
\right)|_{J=0}\; .
$$

By following the same steps as in eqs.(\ref{reprf})-(\ref{gfp3}), we find
\begin{eqnarray}
<\rho({\vec r})> &=&  -{1 \over{T_{eff}}} \; \nabla^2_{\vec
r} \left({{\delta}\over{\delta J({\vec r})}}
  \sum_{N=0}^{\infty}\; {1 \over{N!}}\;
\left [ z\left({m \over{2\pi \beta}}\right)^{3/2}\right]^N
\; \;   {\cal Z}[0]^{-1} \right.\cr \cr
\int\int   \;   {\cal D}\xi & & \left. e^{ -\int d^3x\left[\frac12 \;
(\nabla \xi)^2 
- 2 m  \sqrt{\pi G\beta}\;\xi({\vec x})\;  J({\vec x})\right]}\;
\; \int\ldots \int
\prod_{l=1}^N d^3q_l\; \; e^{ 2 m  \sqrt{\pi G\beta}\; \sum_{l=1}^N
\xi({\vec q}_l)}\right)|_{J=0} \cr\cr
&=&  -{1 \over{T_{eff}}} \; \nabla^2_{\vec
r} \left({{\delta}\over{\delta J({\vec r})}}\;\log {\cal Z}[J(.)]\right)|_{J=0}
 \quad .
\end{eqnarray}

Performing the derivatives in the last formula yields
\begin{equation}
<\rho({\vec r})> = - {1 \over {T_{eff}}}\;   \int\int\; {\cal D}\phi\; \; 
\nabla^2 \phi({\vec r})\;
e^{-{1\over{T_{eff}}}\; \int d^3x \left[ \frac12(\nabla
\phi)^2 \; - \mu^2 \; e^{\phi({\vec x})}\;\right]}\; {\cal Z}[0]^{-1}\; .
\end{equation}
We then see that  in the $\phi$-field language the particle density
expresses as
\begin{equation}\label{rouno}
 \rho({\vec r}) =  -{1 \over {T_{eff}}}\;  \nabla^2 \phi({\vec r}) \; .
\end{equation}

\bigskip

By using eq.(\ref{rouno}), the gravitational potential at the point $ \vec r $ 
$$
U( \vec r ) = -G m \int {{d^3x} \over  { | {\vec x} - {\vec r}|}}\; 
 \rho({\vec x}) \; ,
$$
can  be expressed as
\begin{equation}\label{Ufi}
U( \vec r ) = - {T \over m}\; \phi( \vec r ) \; .
\end{equation}

We can analogously express the correlation functions as
\begin{eqnarray}\label{corre}
C({\vec r_1},{\vec r_2})&\equiv&
<\rho({\vec r_1})\rho({\vec r_2}) > -<\rho({\vec r_1})><\rho({\vec r_2}) > 
\cr \cr
&=&  \left({1 \over{T_{eff}}} \right)^2\; \nabla^2_{\vec r_1}\;
\nabla^2_{\vec r_2}  \;
\left({{\delta}\over{\delta J({\vec r_1})}}\;{{\delta}\over{\delta
J({\vec r_2})}}\; \log{\cal Z}[J(.)]\right)|_{J=0} \; .
\end{eqnarray}
This can be also written as
\begin{equation}\label{corr2}
C({\vec r_1},{\vec r_2}) =  {{\mu^4}\over{T_{eff}}^2} \; \left[  
<e^{\phi({\vec r_1})} \; e^{\phi({\vec r_2})}> -
<e^{\phi({\vec r_1})}> \; <e^{\phi({\vec r_2})}> \right]\; .
\end{equation}

A simple short distance regularization of the Newtonian force for the
two-body potential is
$$
v_a({\vec r}) = -{{G m^2} \over r}\; [ 1 - \theta(a-r) ] \; ,
$$
$ \theta(x)$ being the step function. The cutoff $ a $ can be chosen of
the order of atomic distances but its actual value is unessential.

The introduction of the short-distance cutoff elliminates the unphysical
short distance collapse of the gravitational gas. please notice,
that the interaction between atoms or molecules is repulsive and {\bf not}
gravitational for short distances (Van der Waals forces, for instance).


\subsection{D-dimensional generalization}

This approach generalizes immediately to $D$-dimensional space where
the Hamiltonian (\ref{hami3}) takes then the form
\begin{equation}\label{hamiD}
H_N = \sum_{l=1}^N\;{{p_l^2}\over{2m}} - G \, m^2 \sum_{1\leq l < j\leq N}
{1 \over { |{\vec q}_l - {\vec q}_j|^{D-2}}},\quad  {\rm for}\;  D \neq 2
\end{equation}
and
\begin{equation}\label{hami2}
H_N = \sum_{l=1}^N\;{{p_l^2}\over{2m}} - G \, m^2 \sum_{1\leq l < j\leq N}
\log{1 \over { |{\vec q}_l - {\vec q}_j|}}, \quad  {\rm at}\;  D= 2\; .
\end{equation}

The steps from eq.(\ref{gfp}) to (\ref{zfi}) can be trivially
generalized with the help of the relation
\begin{equation}\label{Dgreen} 
\nabla^2 { 1 \over { | {\vec x} - {\vec y}|^{D-2}}}= -C_D \; \delta( {\vec
x} - {\vec y}) \; 
\end{equation}
in $D$-dimensions and
$$
\nabla^2 \log{ 1 \over { | {\vec x} - {\vec y}|}}= -C_2 \; \delta( {\vec
x} - {\vec y}) \; 
$$
at $ D= 2$. 

Here,
\begin{equation}
C_D \equiv (D-2)\, {{2 \pi^{D/2}}\over {\Gamma(\frac{D}{2})}}\; \; 
{\rm for}~~D\neq 2 \quad {\rm and}~~ C_2 \equiv 2\pi\; .
\end{equation}

We finally obtain as a generalization of  eq.(\ref{zfi}),
\begin{equation}\label{zfiD}
{\cal Z} =  \int\int\;  {\cal D}\phi\;  e^{ -{1\over{T_{eff}}}\;
\int d^Dx \left[ \frac12(\nabla\phi)^2 \; - \mu^2 \; e^{\phi({\vec
x})}\right]}\; , 
\end{equation}

where

\begin{equation}\label{paramD}
\mu^2 =  {{C_D} \over {(2\pi)^{D/2}}}\;
 z\; G \, m^{2+D/2} \, T^{D/2-1} 
\quad , \quad T_{eff} =  C_D \; {{G\; m^2}\over {T}} \; .
\end{equation}

We have then transformed
the partition function for the $D$-dimensional
gas of particles in gravitational interaction into the  partition
function for a 
scalar field $\phi$ with exponential interaction.  
The effective
temperature $ T_{eff} $ for the $\phi$-field partition function is
{\bf inversely} 
proportional to $ T $ for {\bf any} space dimension. The characteristic length
$\mu^{-1}$ behaves as $ \sim T^{-(D-2)/4} $.

\section{Scaling behaviour}

We derive here the scaling behaviour of the $\phi$ field following the
general renormalization group 
arguments in the theory of critical phenomena \cite{kgw,dg}

\subsection{Classical Scale Invariance}

Let us investigate how the  action (\ref{acci}) transforms under scale
transformations 
\begin{equation}\label{trafoS}
{\vec x} \to {\vec x}_{\lambda} \equiv \lambda{\vec x} \; ,
\end{equation}
where $\lambda$ is an arbitrary real number.

In $D$-dimensions the action takes the form
\begin{equation}\label{acciD}
S[\phi(.)] \equiv  {1\over{T_{eff}}}\;
\int d^D x \left[ \frac12(\nabla\phi)^2 \; - \mu^2 \; e^{\phi({\vec
x})}\right] \; .
\end{equation}

We define the scale transformed field $\phi_{\lambda}({\vec x})$ as follows
\begin{equation}\label{filam}
\phi_{\lambda}({\vec x}) \equiv \phi(\lambda{\vec x}) +\log\lambda^2
\; .
\end{equation}
Hence,
$$
(\nabla\phi_{\lambda}({\vec x}))^2 = \lambda^2 \; (\nabla_{x_{\lambda}}
\phi({\vec x}_{\lambda}))^2
\quad , \quad  e^{\phi_{\lambda}({\vec x})}=  \lambda^2  \;
 e^{\phi({\vec x}_{\lambda})}
$$
We find upon changing the integration variable in eq.(\ref{acciD})
from $ {\vec x} $  to  $ {\vec x}_{\lambda} $
\begin{equation}\label{covdil}
S[\phi_{\lambda}(.)] = \lambda^{2-D} \; S[\phi(.)] 
\end{equation}

We thus see that the  action (\ref{acciD})  {\bf scales} under dilatations
in spite of the 
fact that it contains the dimensionful parameter $ \mu^2 $. 
This remarkable scaling property is of course a consequence of the
scale   behaviour of the gravitational interaction in $D$ dimensions 
\cite{prd}.

In particular, in $ D = 2 $ 
the action (\ref{acciD}) is scale invariant. In such
special case, it is moreover conformal invariant.

\bigskip

The (Noether) current associated to the scale transformations 
(\ref{trafoS}) is 
\begin{equation}
J_i({\vec x}) = x_j\; T_{ i j} ({\vec x}) + 2 \;\nabla_i\phi({\vec x})\; ,
\end{equation}
where $  T_{ij} ({\vec x}) $ is the stress tensor
$$
 T_{ i j} ({\vec x}) =  \nabla_i\phi({\vec x}) \; \nabla_j\phi({\vec x})
- \delta_{ij} \; L
$$
and $L \equiv \frac12(\nabla\phi)^2 \; - \mu^2 \; e^{\phi({\vec x})}
$ stands for the action density. That is,
$$
J_i({\vec x}) = ({\vec x}. \nabla\phi + 2)\;  \nabla_i\phi({\vec x})-
x_i \; \left[ \frac12(\nabla\phi)^2 \; - \mu^2 \; e^{\phi({\vec x})}\right]
$$
By using the classical equation of motion (\ref{eqMov}), we then find
$$
 \nabla_i J_i({\vec x}) = (2 - D) L \; .
$$
This non-zero divergence is due to the variation of the action under
dilatations [eq.  (\ref{covdil})].

\bigskip

If $\phi({\vec x})$ is a stationary point of the action (\ref{acciD}):
\begin{equation}\label{eqMov}
\nabla^2\phi({\vec x}) +  \mu^2 \; e^{\phi({\vec x})} = 0 \; ,
\end{equation}
then $ \phi_{\lambda}({\vec x}) $ [defined by eq.(\ref{filam})] is
also a stationary point:
$$
\nabla^2\phi_{\lambda}({\vec x}) +  \mu^2 \; e^{\phi_{\lambda}({\vec
x})} = 0 \; . 
$$

A rotationally invariant stationary point is given by
\begin{equation}\label{fic}
\phi^c(r) = \log{{2(D-2)}\over { \mu^2 r^2}} \; .
\end{equation}
This singular solution is {\bf invariant} under the scale
transformations (\ref{filam}). That is
$$
\phi^c_{\lambda}(r) =\phi^c(r) \; .
$$
Eq.(\ref{fic}) is dilatation and rotation invariant. 
It provides the {\bf most symmetric} stationary point of
the action. Notice that there are no constant stationary solutions
besides the singular solution $ \phi_0 = -\infty $.

\subsection{Thermal Fluctuations}

In this section we compute the partition function eqs.(\ref{zfi}) and
(\ref{zfiJ}) by saddle point methods.

Eq.(\ref{eqMov}) admits  only one constant stationary solution
\begin{equation}\label{fiS}
 \phi_0 = -\infty \; .
\end{equation}

In order to make such solution finite we now introduce a
regularization term $ \; \epsilon \, \mu^2 \, \phi({\vec x}) $ with $
\epsilon << 1 $ in the action  $ S $ [eq.(\ref{acci})]. This
corresponds to an action density
\begin{equation}\label{densac}
L =  \frac12(\nabla\phi)^2 \; + \; u(\phi)
\end{equation}
where
$$
 u(\phi) =  - \mu^2 \; e^{\phi({\vec x})} +  \epsilon \; \mu^2 \;
 \phi({\vec x}) \; .
$$
This extra term can be obtained by adding a small constant term $
-\epsilon \; \mu^2/T_{eff} $ to $\rho({\vec x})$ in eqs.(\ref{defro}) -
(\ref{reprf}). This a simple way to make $ \phi_0 $ finite.

We get in this way a constant stationary point at $   \phi_0 =
\log\epsilon $ where $ u'(\phi_0) = 0 $. However, scale invariance is
broken since $  u''(\phi_0) = - \epsilon \; \mu^2 \neq 0 $. We can add
a second regularization term to   $ \; \frac12 \, \delta \, \mu^2 \;
 \phi({\vec x})^2 \; $  to $ L $, (with $  \delta << 1 $) in order to enforce $
u''(\phi_0) = 0 $. This quadratic term amounts to a long-range
shielding of the gravitational force. 
We finally set:
$$
 u(\phi) =  - \mu^2 \left[  e^{\phi({\vec x})} -  \epsilon \;
 \phi({\vec x}) -  \frac12 \; \delta \; \phi({\vec x})^2 \right]
 \; ,
$$
where the two regularization parameters $ \epsilon$ and $ \delta $ are
related by 
$$ 
 \epsilon( \delta ) = \delta [1 - \log \delta] \; ,
$$
and the stationary point has the value
$$
  \phi_0 =\log\delta \; .
$$

Expanding around $ \phi_0 $ 
$$
\phi({\vec x}) =  \phi_0 + g \; \chi({\vec x})
$$
where $ g \equiv \sqrt{\mu^{D-2} \, T_{eff}} $ and 
$ \chi({\vec x}) $ is the fluctuation field, yields
\begin{equation}\label{fiinfi}
\frac{1}{g^2}\;  L =  \frac12 \; (\nabla\chi)^2 \;
- {{\mu^2 \delta}\over {g^2}} \left[ e^{g \chi} -1 - g \; \chi - 
 \frac12 \; g^2 \;  \chi^2 \right]
\end{equation}
We see
perturbatively in $g$ that  $ \chi({\vec x}) $ is a {\bf massless} field.

\bigskip

Concerning the boundary conditions, for the ISM 
we must consider the system inside
a large sphere of radius $R \; ( 10^{-4}\; - \; 10^{-2}
\;  pc \;   \leq     R   \leq 100\;  pc )$. That is, all
integrals are computed over such large sphere.

\bigskip

Using eq.(\ref{rouno}) the particle density takes now the form
$$
 \rho({\vec r}) =  -{1 \over {T_{eff}}}\;  \nabla^2 \phi({\vec r}) = 
 -{g \over {T_{eff}}}\;  \nabla^2 \chi({\vec r}) =  {{\mu^2
 \delta}\over { T_{eff} }} \left[ e^{g \chi({\vec r})} -1 - g
 \chi({\vec r})  \right]   \; .
$$
It is convenient to renormalize the  particle density 
by its stationary value $ \delta = e^{\phi_0} $,
\begin{equation}\label{renro}
 \rho({\vec r})_{ren} \equiv \frac{1}{\delta} \;  \rho({\vec r}) =
 {{\mu^D}\over {g^2 }} \left[ e^{g \chi({\vec r})} -1 - g \chi({\vec
 r})  \right]   \; .
\end{equation}
We see that in the $ \delta \to 0 $ limit the interaction  in 
eq.(\ref{fiinfi}) vanishes. No infrared divergences  appear in the
Feynman graphs calculations,  since we work on a
very large but finite volume of size $ R $. Hence, in the  $\delta \to
0 $ limit, the whole perturbation series around $ \phi_0 $  reduces  to the
zeroth order term.

The constant saddle point $ \phi_0 $ fails to catch up  the whole field
theory content. In fact, more information arises perturbing around the
stationary point $ \phi^c(r) $ given by eq.(\ref{fic}) \cite{fut}. 

Using eqs.(\ref{corr2}),  (\ref{fiinfi}) and
(\ref{renro}) we obtain for the density correlator in the   $ \delta
\to 0 $ limit, 
$$
C({\vec r_1},{\vec r_2}) =   {{\mu^{2 D}}\over {g^4}}\;
\left\{ \exp\!\left[{{g^2}\over { C_D\;  \, 
  \left( \mu \, | {\vec  r_1} - {\vec  r_2}|\right)^{D-2}}} \right] -1 
- {{g^2}\over {  C_D \;    \left(  \mu \, | {\vec  r_1} - {\vec
r_2}|\right)^{D-2}}} \right\} \; .
$$

For large distances, we find

\begin{equation}\label{corrasi}
 C({\vec r_1},{\vec r_2}) \buildrel{  | {\vec  r_1} - {\vec  r_2}|\to
\infty}\over =  {{ \mu^4 }\over {2\, C_D^2 \; 
 | {\vec  r_1} - {\vec  r_2}|^{2(D-2)}}} + O\left( \; | {\vec
r_1} - {\vec r_2}|^{-3(D-2)}\right)\; . 
\end{equation}

That is, the  $\phi$-field theory  {\bf scales}.  Namely, the
theory behaves  critically for a {\bf continuum set} of  values of $\mu$ and
$ T_{eff} $.  

Notice that the
density correlator $C({\vec r_1},{\vec r_2})$ behaves for large
distances as the correlator of $ \chi({\vec r})^2 $. This stems from
the fact that $ \chi({\vec r})^2 $ is the most relevant operator in
the series expansion of the density (\ref{renro})
\begin{equation}\label{denser}
 \rho({\vec r})_{ren} = \frac12 \; \mu^D \;  \chi({\vec r})^2 + O(\chi^3)\; .
\end{equation}

As remarked above, the constant stationary point $ \phi_0 = \log\delta \to
-\infty $ only produces  the zeroth order of perturbation theory.
More information arises perturbing around the stationary point $ \phi^c(r) $
given by eq.(\ref{fic}) \cite{fut}. 

We succeded in this way to to construct a scale invariant 
correlations in perturbation theory for the expenential interaction. 
The existence of such expansion is a necessary
condition to show that the theory is scale invariant. Perturbation theory
around the stationary point (\ref{fic})
shares such scale invariant property \cite{fut}.

\subsection{Renormalization Group and Finite Size Scaling Analysis}

As is well known \cite{kgw,dg,nn}, physical magnitudes  for {\bf
infinite} volume systems diverge at the critical point  as $ \Lambda $ to a
negative power.  $ \Lambda $ measures the distance to the critical
point. (In condensed matter and spin systems, $ \Lambda $ is
proportional to the temperature minus the critical temperature \cite{dg,nn}).
One has  for the correlation length  $ \xi $, 
$$ 
\xi( \Lambda ) \sim  \Lambda^{-\nu} \; ,
$$
and for the specific heat $ {\cal C} $,
\begin{equation}\label{calor}
 {\cal C} \sim  \Lambda^{-\alpha}  \; .
\end{equation}
Correlation functions scale at criticality. For example,  the 
scalar field $\phi$ (which in spin systems describes the magnetization),
$$
<\phi({\vec r})\phi(0)> \sim r^{-1-\eta} \; .
$$

The critical exponents $\nu, \;\alpha $ and $ \eta $ are pure numbers
that depend only on the universality class  \cite{kgw,dg,nn}.

For a {\bf finite} volume system, all physical  magnitudes are  {\bf
finite} at the critical point. Indeed, for a  system whose  size $ R $
is large, the  physical  magnitudes
take large values at the critical point. Thus, for large   $ R $, one can
use the infinite volume theory to treat finite size systems at
criticality. In particular,  the correlation length provides the
relevant physical length $ \xi \sim R $. This implies that
\begin{equation}\label{fss}
\Lambda \sim R^{-1/\nu} \; .
\end{equation}
We can apply these concepts to the  $\phi$-theory 
since, as we have seen in the previous section, it 
exhibits scaling in a finite volume $\sim R^3 $. 
Namely, the two points correlation function exhibits a power-like
behaviour in perturbation theory as shown by  eq.(\ref{corrasi}). This happens 
 for a  {\bf continuum set} of  values of 
$T_{eff}$ and $\mu^2$. Therefore, changing $\mu^2/T_{eff}$ keeps the
theory in the scaling region. 
At the point $ \mu^2/T_{eff} = 0 $, the partition function $ {\cal Z}
$ is singular. From eq.(\ref{muyT}),  we shall thus identify
\begin{equation}\label{zcritico}
  \Lambda \equiv   {{\mu^2}\over{T_{eff}}} = z\,
  \left({{mT}\over{2\pi}}\right)^{3/2} \; .
\end{equation}
Notice that the critical point $  \Lambda = 0 $, corresponds to zero
fugacity. [For simplicity, we restrict from now on to the three
dimensional case]. 

Thus,  the partition function in the scaling regime can be written as 
\begin{equation}\label{Zsca1}
{\cal Z}(\Lambda) = 
 \int\int\;  {\cal D}\phi\;  e^{ -S^* + \Lambda
\int d^3x  \; e^{\phi({\vec x})}\;}\; ,
\end{equation}
where $S^*$ stands for the action (\ref{acci}) at the critical point
$\Lambda = 0 $.

We define the renormalized mass  density  as
\begin{equation}\label {dfensi}
m\, \rho({\vec x})_{ren} \equiv m\, \,  e^{\phi({\vec x})}
\end{equation}
and we identify it with the  energy density in the renormalization
group. [Also called the `thermal perturbation operator'.]
This identification  follows from the fact that they are the most
relevant positive definite operators. Moreover, such  identification is
supported by the perturbative result (\ref{denser}).

In the scaling regime we have \cite{dg} for the logarithm of the
partition function
\begin{equation}\label{Zsca2}
{1 \over V} \; \log{\cal Z}(\Lambda) = {K \over{(2-\alpha)(1-\alpha)}}\;
\Lambda^{2-\alpha} + F(\Lambda) \; ,
\end{equation}
where   $  F(\Lambda) $ is an analytic
function of $ \Lambda $ around the origin 
$$ 
F(\Lambda) = F_0 + a \; \Lambda + \frac12 \, b  \; \Lambda^2 + \ldots \; .
$$ 
 $ V = R^D $ stands for the volume and $ F_0, \; K, \; a $ and $ b $
are constants. 

Calculating the logarithmic derivative of ${\cal Z}(\Lambda)$ with
respect to $ \Lambda $ from eqs.(\ref{Zsca1}) and from (\ref{Zsca2})
and equating the results yields
\begin{equation}\label{masaR}
{1 \over V} \;{{\partial}\over{\partial\Lambda}}\log{\cal Z}(\Lambda)=
a +  {K \over{1-\alpha}}\,
\Lambda^{1-\alpha}
+ \ldots = {1 \over V} \int d^Dx  \; <e^{\phi({\vec x})}>\; .
\end{equation}
where we used  the scaling
relation $ \alpha = 2 - \nu D $ \cite{dg,nn}.

We can apply here finite size scaling arguments and 
 replace $\Lambda$ by $\sim R^{-\frac{1}{\nu}}$ [eq.(\ref{fss})],
$$
{{\partial}\over{\partial\Lambda}}\log{\cal Z}(\Lambda)= V \, a +
 {K \over{1-\alpha}}\, R^{1/\nu} + \ldots\; . 
$$

Recalling eq.(\ref{dfensi}), we can express the mass contained in a region
of size $ R $ as
\begin{equation}\label{defM}
M(R) = m  \int^R e^{\phi({\vec x})} \; d^Dx  \; .
\end{equation}
Using  eq.(\ref{masaR}) we find
$$
<M(R)> =  m  \, V \, a +m  \, {K \over{1-\alpha}}\; R^{ \frac1{\nu}} +
\ldots\; . 
$$
and
\begin{equation}\label{Sdensi}
<\rho({\vec r})> = m  a \; +m \, {K
\over{\nu(1-\alpha)\Omega_D}}\;r^{ \frac1{\nu}-D} \quad {\rm for}\; r
\;  {\rm of ~ order}\;\sim R .
\end{equation}
where $ \Omega_D $ is the surface of the unit sphere in $D$-dimensions.

The energy density correlators obey the
renormalization group equations at criticality \cite{dg} -\cite{nn}. 
One has for the connected $L$-point correlations of the energy density 
(`thermal operator'),
\begin{equation}\label{GRcorr}
\left[ \sum_{k=1}^L {\vec x}_k . {{\partial}\over {\partial {\vec x}_k}} +
L\, (3 - \frac1{\nu} ) \right]
<\rho({\vec x_1}) \ldots \rho({\vec x_L}) >^{conn} = 0 \; ,
\end{equation}
for the  three dimensional space.
The homogeneity assumption implies that 
$ <\rho({\vec x_1})\rho({\vec x_2}) > $ 
is only a function of the difference 
$ {\vec x_1} - {\vec x_2} $ for large $ | {\vec x_1} - {\vec x_2} | $. 
We can therefore write for the density-density correlators
(\ref{corre}) in $ 3 $ space dimensions
using by eq.(\ref{GRcorr}) for $ L = 2 $,
\begin{equation}\label{corrG}
C({\vec r_1},{\vec r_2})\sim |{\vec r_1} -{\vec r_2}|^{\frac2{\nu} -6} \; .
\end{equation}
where both $ {\vec r_1} $ and $ {\vec r_2} $ are  inside the
finite  volume $ \sim R^D $.

The perturbative calculation (\ref{corrasi})  matches
with this result for $ \nu = \frac12 $. That is, the mean field
value for the exponent $ \nu $. 

Let us now compute the second derivative of $ \log{\cal Z}(\Lambda) $
with respect to $\Lambda$ in two ways.  We find from eq.(\ref{Zsca2}) 
$$
{{\partial^2}\over{\partial\Lambda^2}}\log{\cal Z}(\Lambda)= V\left[
\Lambda^{-\alpha} \, K + b  + \ldots  \right] \; .
$$
We get from eq.(\ref{Zsca1}),
\begin{equation}\label{flucM}
{{\partial^2}\over{\partial\Lambda^2}}\log{\cal Z}(\Lambda)=
\int d^Dx\; d^Dy\; C({\vec x},{\vec y}) \sim R^D \int^R 
{{ d^3x}\over{x^{2D - 2d_H}}}  \sim \Lambda^{-2}\sim R^D \;
\Lambda^{-\alpha} 
\end{equation}
where we used eq.(\ref{fss}), eq.(\ref{corrG}) and the scaling
relation $ \alpha = 2 - \nu D $ \cite{dg,nn}. We
conclude that the scaling 
behaviours,   eq.(\ref{Zsca2})  for the partition function, 
eq.(\ref{calor}) for the specific heat and  eq.(\ref{corrG}) for the
two points correlator are consistent.
In addition,   eqs.(\ref{defM}) and (\ref{flucM})
yield for the  mass fluctuations  squared
$$
(\Delta M(R))^2 \equiv  \; <M^2> -<M>^2  \; \sim
\int d^Dx\; d^Dy\; C({\vec x},{\vec y}) \sim R^{2d_H}\; .
$$
Hence, 
\begin{equation}\label{Msca}
\Delta M(R)  \sim  R^{d_H}\; .
\end{equation}

\bigskip

The scaling exponent $\nu$ can be identified with the inverse
Haussdorf (fractal) dimension $d_H$ of the system
$$
d_H = \frac1{\nu} \; .
$$
In this way, $ M \sim R^{d_H} $ according to the usual  definition of fractal
dimensions \cite{sta}.

\medskip

The Fourier transform,
$$
{\tilde \rho}({\vec k}) \equiv \int d^3r \; e^{i{\vec k}.{\vec r}}\;
{\rho}({\vec r}) \; ,
$$
will scale as:
\begin{equation}
<{\tilde \rho}({\vec k})> \sim k^{ -\frac1{\nu}} \; .
\end{equation}

This quantity is related to the number of clouds with mass $ M $ ,
$ n(M) $ , as follows
\begin{equation}\label{nM}
 n(M) \; \sim  \;<{\tilde \rho}(M)>  \;\sim \; M^{-x}\; .
\end{equation}

From eqs.(\ref{fou}) and (\ref{nM}) we find
$$
x = d_H \; ,
$$
which is in agreement with \cite{lar2}.

Using eq.(\ref{corrG})  we can compute the average potential energy as
$$
< {\cal V}> =   \frac12 \,\beta \, G \, m^2
\int_{ | {\vec x} - {\vec y}|> a }^R \;
{{d^3x\, d^3y}\over { | {\vec x} - {\vec y}|}}\;  C({\vec x},{\vec y})
\sim R^{\frac2{\nu} -1} \; .
$$

From here and eq.(\ref{Msca})
we get as  virial estimate for the atoms kinetic energy
\begin{equation}
<v^2> = {{< {\cal V}>}\over {M(R)}} \sim  R^{\frac1{\nu} -1} \; .
\end{equation}
This corresponds to a velocity dispersion
\begin{equation}\label{Vsca}
\Delta v \sim R^{\frac12(\frac1{\nu} -1)} \; .
\end{equation}
That is,  we predict [see eq.(\ref{vobser})] a new scaling relation
$$
q =\frac12\left(\frac1{\nu} -1\right) =\frac12(d_H -1)  \; .
$$

\bigskip

The calculation of the critical amplitudes [that is, the coefficients in 
front of the powers of $ R $ in eqs.(\ref{Msca}), (\ref{corrG}) and
 (\ref{Vsca}) ] is beyond the scope of the present paper \cite{fut}.

The $\phi$-field connected correlators obey at criticality the
renormalization group equation \cite{dg} -\cite{nn}

\begin{equation}\label{fiGRcorr}
\left[ \sum_{k=1}^N {\vec x}_k . {{\partial}\over {\partial {\vec x}_k}} +
\frac{N}{2}\, (1+ \eta ) \right]
<\phi({\vec x_1}) \ldots \phi({\vec x_L}) >^{conn} = 0 \; ,
\end{equation}
for the  three dimensional space.

It follows from eq.(\ref{fiGRcorr}) that the
two-point subtracted correlator of the gravitational potential  behaves
 for long distances as
$$
< \phi({\vec x}_1) \phi( {\vec x}_2) > \; \sim  |{\vec x}_1 - {\vec
x}_2|^{-1-\eta} \; .
$$
\subsection{Values of the scaling exponents and the fractal dimensions}

The scaling exponents $ \nu , \; \alpha $ considered in sec IIIC can
be computed through the renormalization group approach. The case of a 
 {\bf single} component scalar field has been extensively studied 
in the literature \cite{dg,nn,grexa}. Very probably, there is an
unique, infrared stable fixed point in three space dimensions: the
Ising  model fixed point. Such  non-perturbative fixed point is
reached in the long scale regime independently of the initial shape of
the interaction $ u(\phi) $ [eq.(\ref{densac})]  \cite{grexa}.

The numerical values of the scaling exponents associated to the 
Ising  model fixed point are

\begin{equation}\label{Ising}
\nu = 0.631...  \quad , \quad d_H = 1.585...  \quad {\rm and} \quad
\alpha = 0.107... \; \; .
\end{equation}

\bigskip

In the $\phi$ field model there are two dimensionful parameters: $\mu$
and $T_{eff}$. The dimensionless combination 
$$
g^2 = \mu \, T_{eff} =  (8 \pi)^{3/4}\; \sqrt{z} \; \; {{G^{3/2}\;
m^{15/4}}\over T^{3/4}} 
$$
acts as the coupling constant for the non-linear
fluctuations of the field $\phi$. 

Let us consider a gas formed by neutral hydrogen at thermal equilibrium with
the cosmic microwave background. We set $ T = 2.73\, K $ and  estimate
the fugacity $ z $ using the ideal gas value
$$
z = \left( {{2\pi}\over {m T } }\right)^{3/2}\; \rho \; .
$$
Here we use $  \rho = \delta_0 $ atoms cm$^{-3}$ for the ISM density and
$  \delta_0 \simeq 10^{10} $. Eq. eqs.(\ref{muyT}) yields
\begin{equation}\label{valN}
{1 \over {\mu }} = 2.7 \; {1 \over { \sqrt{\delta_0}}}\; {\rm  AU} \sim
30 \; {\rm AU} \quad {\rm and} \quad 
g^2 = \mu \, T_{eff} = 4.9 \; 10^{-58} \; \sqrt{\delta_0} \sim 5 \,10^{-53}
 \; . 
\end{equation}

This extremely low value for $g^2 $ 
suggests that the perturbative calculation [sec. IIIB] may apply here 
yielding the mean field values for the exponents, i. e.  
\begin{equation}\label{campM}
\nu = 1/2 \quad ,  \quad d_H = 2   \quad , \quad \eta =0 \quad {\rm
and } \quad \alpha = 0 \; .
\end{equation}
That is, the effective coupling constant grows with
the scale according to the renormalization group flow (towards the
Ising fixed point). Now, if the extremely low value of the initial
coupling eq.(\ref{valN}) applies, the perturbative result (mean field)
will hold for many scales (the effective $ g $ grows roughly as the
length).

$ \mu^{-1} $ indicates the order of the smallest distance where the
 scaling regime applies.  A safe
lower bound supported by observations is around $20$ AU $\sim 3.\,
10^{14}$ cm , in agreement with our estimate.

\bigskip

Our theoretical predictions for $ \Delta M(R) $ and $ \Delta v $
[eqs.(\ref{Msca}) and (\ref{Vsca})] both for 
the Ising  eq.(\ref{Ising}) and for the mean field values
eq.(\ref{campM}), are in  agreement with the astronomical
observations [eq.(\ref{vobser})]. The present observational bounds on
the data are  larger than the difference between the mean field and
Ising values of  the exponents $ d_H $ and $ q $. 

Further theoretical work in the $\phi$-theory will determine whether
the scaling behaviour is given by the mean field or by the Ising fixed
point \cite{fut}.

\subsection{The two dimensional gas and random surfaces fractal dimensions} 

In the two dimensional case ($D=2$) the partition function
(\ref{zfiD}) describes  the Liouville model that arises in string
theory\cite{poly} and in the theory of random surfaces 
(also called two-dimensional quantum gravity).
For strings in $c$-dimensional Euclidean space the  partition function
takes the form\cite{poly}
\begin{equation}\label{zfiL}
{\cal Z}_c =  \int\int\;  {\cal D}\phi\;  e^{ -{{26-c}\over{24\pi}}\;
\int d^2x \left[ \frac12(\nabla\phi)^2 \; + \mu^2 \; e^{\phi({\vec
x})}\right]}\; . 
\end{equation}
This coincides with eq.(\ref{zfiD}) at $D=2$ provided we flip the sign
of $ \mu^2 $ and identify the parameters (\ref{paramD}) as follows,
\begin{equation}
T = G m^2\; {{26-c}\over 12}\quad , \quad \mu^2 = z G m^3 \; .
\end{equation}

Ref.\cite{amb} states that $ d_H = 4 $ for $ c \leq 1 $, $ d_H = 3 $ for
$ c = 2 $ and  $d_H = 2$ for
$ c \geq 4 $. In our context this means
$$
d_H =2 \;\; {\rm for}~~ T \leq    \frac{25}{12} \;  G m^2 \quad ,    \quad
 d_H = 3 \;\; {\rm for}~~ T = 2\, G m^2  \quad
{\rm and} \quad
 d_H =4 \;\; {\rm for}~~ T \geq \frac{11}6  \;  G m^2 \; .
$$

For $ c \to \infty , \; g^2 \to 0 $ and we can use the perturbative
result (\ref{corrasi}) yielding $ \nu = \frac12 , \; d_H = 2 $
in agreement with the above discussion for $ c \geq 4 $. 

\subsection{Stationary points, the Poisson equation and the Jeans length}

The stationary points of the $\phi$-field partition function (\ref{zfi}) 
are given by the non-linear partial differential equation
$$
\nabla^2\phi = -\mu^2\,  e^{\phi({\vec x})} \; .
$$
In terms of the gravitational potential $U({\vec x})$ [see eq. (\ref{Ufi})],
this takes the form

\begin{equation}\label{equih}
\nabla^2U({\vec r}) = 4 \pi G \, z \,  m
\left({{mT}\over{2\pi}}\right)^{3/2} \,  e^{ - \frac{m}{T}\,U({\vec r})} \; .
\end{equation}
This corresponds to the Poisson equation for a thermal matter distribution
fulfilling an ideal gas in hydrostatic equilibrium,
as can be seen as follows \cite{sas}.
The hydrostatic equilibrium condition 
$$
\nabla P({\vec r}) = - m \, \rho({\vec r}) \; \nabla U({\vec r})\; ,
$$
where $ P({\vec r}) $ stands for the pressure, combined with the
equation of state for the ideal gas
$$
P = T \rho \; ,
$$
yields for the particle density
$$
 \rho({\vec r}) =  \rho_0 \; e^{ - \frac{m}{T}\,U({\vec r})} \; ,
$$
where $ \rho_0 $ is a constant. Inserting this relation into the
Poisson equation 
$$
\nabla^2U({\vec r}) = 4 \pi G\, m \, \rho({\vec r})
$$
yields eq.(\ref{equih}) with 
\begin{equation} \label{RO0}
  \rho_0 =  z \,\left({{mT}\over{2\pi}}
\right)^{3/2} \; . 
\end{equation}

For large $ r $,  eq.(\ref{equih}) gives a density decaying as $
r^{-2} $ ,
\begin{equation}
 \rho({\vec r}) \buildrel{r\to \infty}\over = {T\over{2\pi G m}}\,
\frac1{r^2} \,\left[ 1 + O\left(\frac1{\sqrt{r}} \right)  \right]
\quad , \quad U({\vec r}) \buildrel{r\to \infty}\over = 
\frac{T}{m}\;\log\left[{{2\pi G  \rho_0}\over T}\; r^2\right] +
O\left(\frac1{\sqrt{r}} \right) \; .
\end{equation}
Notice that this density, which describes a single stationary solution, 
decays for large $r$ {\bf faster} than the density (\ref{Sdensi}) governed by
thermal fluctuations.

\bigskip

Spherically symmetric solutions of eq.(\ref{equih}) has been studied 
in detail \cite{chandra}.
The small fluctuations around such isothermal spherical solutions
as well as the stability problem were studied in \cite{kh}.

\bigskip

The Jeans distance is in this context,
\begin{equation}\label{distaJ}
d_J \equiv \sqrt{ 3 T \over m}\; {1 \over{\sqrt{G\, m \, \rho_0}}} = 
{{ \sqrt{ 3}\; (2\pi)^{3/4}}\over{  \sqrt{z\, G}\; m^{7/4}\; T^{1/4}}}
\; .
\end{equation}
This distance precisely coincides with $ \mu^{-1} $ [see eq.(\ref{muyT})] up to
an inessential numerical coefficient ($\sqrt{12/\pi}$). Hence,  $
\mu $, the only dimensionful parameter in the $\phi$-theory can
be interpreted as the inverse of the Jeans distance.

We want to notice that in the critical regime,  $ d_J $ grows as
$$
d_J \sim R^{d_H/2} \; ,
$$
since $  \rho_0 = \Lambda \sim R^{-d_H} $ vanishes as 
can be seen from  eqs.(\ref{fss}), (\ref{zcritico}) and
(\ref{RO0}). In this estimate  we should use for consistency
the mean field value $ d_H = 2 $, which yields  $ d_J \sim R$.

This  shows that the Jeans distance is  of the  order of the {\bf size} of the
system. The  Jeans distance   {\bf scales} and the instability is
therefore present for all sizes $ R $.

Had  $ d_J $ being   of order larger than $ R $, the Jeans instability
would be absent. 

The fact that  the Jeans instability is  present  {\bf precisely}
at $  d_J \sim R $ is probably essential to  the scaling regime and to
the self-similar (fractal) structure of the gravitational gas. 

\section{Galaxy Distributions}

One obvious feature of galaxy and cluster distributions in the sky is their
hierarchical property: galaxies gather in groups, that are embedded in
clusters, then in superclusters, and so on. (Shapley 1934, Abell 1958). 
Moreover, galaxies and clusters appear to obey scaling properties,
such as the power-law of the two point-correlation function:
$$
\xi(r) \propto r^{-\gamma}
$$
with the slope $\gamma$, the same for galaxies and clusters, of $\approx$ 1.7
(e.g. \cite{peeb}).  This scale-invariance has suggested very
early the idea of fractal models for the clustering hierachy of galaxies
(de Vaucouleurs 1960, 1970; Mandelbrot 1975). Since then, many authors have
shown that a fractal distribution indeed reproduces quite well the 
aspect of galaxy catalogs, for example by simulating a fractal and observing
it, as with a telescope (Scott, Shane \& Swanson, 1954; Soneira \& Peebles 
1978). Sometimes the analysis has been done in terms of a multi-fractal
medium (Balian \& Schaeffer 1989, Castagnoli \& Provenzale 1991, 
Martinez et al 1993, Dubrulle \& Lachieze-Rey 1994).

There is some ambiguity in the definition of the two-point correlation
function $\xi(r)$ above, since it depends on the assumed scale
beyond which the universe is homogeneous; indeed it includes a normalisation by
the average density of the universe, which, if the homogeneity scale is not 
reached, depends on the size of the galaxy sample.
Once $\xi(r)$ is defined, one can always determine a  length $r_0$ where 
$\xi(r_0)$ =1 (Davis \& Peebles 1983, Hamilton 1993).  For galaxies,
the most frequently reported value is $r_0 \approx 5 h^{-1}$ Mpc
(where $h = H_0$/100km s$^{-1}$Mpc$^{-1}$), but it has been shown to increase 
with the distance limits of galaxy catalogs (Davis et al 1988).
$r_0$ is called `correlation length' in the galaxy literature. 
[The notion of correlation length $\xi_0$ is usually different in physics,
where  $\xi_0$ characterizes the exponential decay of correlations $ (\sim 
e^{- r/ \xi_0} ) $. For power decaying correlations, it is said that the  
correlation length is infinite].

The same problem occurs for the two-point correlation function of
galaxy clusters; the corresponding $\xi(r)$ has the same power law 
as galaxies, their  length  $r_0$ has been reported to be about 
$r_0 \approx 25 h^{-1}$ Mpc, and their correlation amplitude is therefore
about 15 times higher than that of galaxies
(Postman, Geller \& Huchra 1986, Postman, Huchra \& Geller 1992).
The latter is difficult to understand, unless there is a considerable
difference between galaxies belonging to clusters and field galaxies (or
morphological segregation). The other obvious explanation is that
the normalizing average density of the universe was then chosen lower.

This statistical analysis of the galaxy catalogs has been criticized by
Pietronero (1987), Einasto (1989) and Coleman \& Pietronero (1992), 
who stress the unconfortable dependence of $\xi(r)$ and of the  length $r_0$
upon the finite size of the catalogs, and on the {\it a priori} assumed 
value of the large-scale homogeneity cut-off.  A way to circumvent these 
problems is to deal instead with the average density as a function of size 
(cf \S 2). It has been shown that the galaxy distribution behaves as a pure
self-similar fractal over scales up to $\approx 100 h^{-1}$ Mpc,
the deepest scale to which the data are statistically robust
(Sylos Labini et al 1996; Sylos Labini \& Pietronero 1996).
This is more consistent with the observation of contrasted large-scale
structures, such as superclusters, large voids or great walls of
galaxies of $\approx 200 h^{-1}$ Mpc (de Lapparent et al 1986, Geller 
\& Huchra 1989). After a proper statistical analysis of all available
catalogs (CfA, SSRS, IRAS, APM, LEDA, etc.. for galaxies, and Abell and
ACO for clusters) Pietronero et al (1997) state that the 
transition to homogeneity might not yet have been reached up to the deepest
scales probed until now. At best, this point is quite controversial,
and the large-scale homogeneity transition is not yet well known. 

Isotropy and homogeneity are expected at very large scales from the
Cosmological Principle (e.g. Peebles 1993). However, this does not imply
local or mid-scale homogeneity (e.g. Mandelbrot 1982, Sylos Labini 1994):
a fractal structure can be locally isotropic, but inhomogeneous.
The main observational evidence in favor of the Cosmological Principle
is the remarkable isotropy of the
cosmic background radiation (e.g. Smoot et al 1992), that provides information
about the Universe at the matter/radiation decoupling. There must therefore
exist a transition between the small-scale fractality to large-scale
homogeneity. This transition is certainly smooth, and might correspond to the
transition from linear perturbations to the non-linear gravitational collapse 
of structures. The present catalogs do not yet see the transition since
they do not look up sufficiently back in time. It can be noticed that
some recent surveys begin to see a different power-law
behavior at large scales ($\lambda \approx 200-400  h^{-1}$ Mpc, e.g. 
Lin et al 1996).

\medskip

There are several approaches to understand non-linear clustering, and therefore
the distribution of galaxies, in an infinite gravitating system. 
Numerical simulations have been widely used, in the hope to trace back from 
the observations the initial mass spectrum of fluctuations, and to test 
postulated cosmologies such as CDM and related variants (cf Ostriker 1993). 
This approach 
has not yet yielded definite results, especially since the physics of
the multiple-phase universe is not well known. Also numerical limitations
(restricted dynamical range due to the softening and limited volume) have
often masked the expected self-similar behavior (Colombi et al 1996).
 A second approach, which 
should work essentially in the linear  (or weakly non-linear) regime, 
is to solve the BBGKY hierarchy
through closure assumptions (Davis \& Peebles 1977; Balian \& Schaeffer 1989).
The main assumption is that the $N$-points 
correlation functions are
scale-invariant and behave as power-laws like is observed for the few-body
correlation functions. Crucial to this approach is the determination of
the void probability, which is a series expansion of the 
$N$-points correlation 
functions (White 1979). The hierachical solutions found in this frame
agree well with the simulations, and with the fractal structure of the universe
at small-scales (Balian \& Schaeffer 1988). 
A third approach is the thermodynamics of gravitating systems, developped
by Saslaw \& Hamilton (1984), which assumes quasi thermodynamic
equilibrium. The latter is justified at the small-scales of non-linear
clustering, since the expansion time-scale is slow with respect to local
relaxation times. Indeed the main effect of expansion is to subtract 
the mean gravitational field, which is negligible for structures of
mean densities several orders of magnitude above average.
  The predictions of the thermodynamical theory have been successfully
compared with N-body simulations (Itoh et al 1993), but a special physical
parameter (the ratio of gravitational correlation energy to thermal energy) 
had to be adjusted for a better fit (Bouchet et al 1991, Sheth \& Saslaw 1996,
Saslaw \& Fang 1996). 

We present in \cite{gal}  a new approach based on field theory and the 
renormalisation group to understand the clustering
behaviour of a self-gravitating expanding universe. We also consider
the thermodynamics properties of the system, assuming quasi-equilibrium
for the range of scales concerned with the non-linear regime and
virialisation. We find an exact mapping between the  self-gravitating gas 
and a continuous field theory for a
single scalar field with an exponential self-coupling. 
This allows us to use  statistical field theory
and the renormalisation group to determine the scaling behaviour.
The small-scale fractal universe can be considered critical with large density
fluctuations developing at any scale. We derive
the corresponding critical exponents. They are very close
to those measured on galaxy catalogs through statistical
methods based on the average density as function of size; these methods
reveal in particular a fractal dimension $D \approx 1.5-2$ 
(Di Nella et al 1996, Sylos Labini \& Amendola 1996, Sylos Labini et al 1996).
This fractal dimension is strikingly close to that observed for
the interstellar medium or ISM (e.g. Larson 1981, Falgarone et al 1991).
We show in ref.\cite{gal} that the theoretical framework based on 
self-gravity that we have developped for the ISM (de Vega, S\'anchez \& Combes
1996a,b, hereafter dVSC) is also  a dynamical mechanism 
leading to the small scale fractal structure of the universe.
This theory is powerfully predictive without any free parameter. 
It allows to compute the $N$-points density correlations without any extra 
assumption.

\section{Galaxy Correlation functions and mass density in a fractal}

The use of the two point correlation function $\xi(r)$ widely spread
in galaxy distributions studies, is based on the assumption that the
Universe reaches homogeneity on a scale smaller than the sample size.
It has been shown by Coleman, Pietronero \& Sanders (1988) 
and Coleman \& Pietronero (1992) that such an hypothesis could
perturb significantly the results.
The correlation function is defined as
$$
\xi(r) = \frac{<n(r_i).n(r_i+r)>}{<n>^2} -1
$$
where $n(r)$ is the number density of galaxies, and $<...>$ is the volume
average (over $d^3r_i$). The  length $r_0$ is defined
by $\xi(r_0) = 1$. The function $\xi(r)$ has a power-law behaviour 
of slope $-\gamma$ for $r< r_0$, then it turns down to zero 
rather quickly at the statitistical limit of the sample. This rapid
fall leads to an over-estimate of the small-scale $\gamma$.
Pietronero (1987) introduces the conditional density
$$
\Gamma(r) = \frac{<n(r_i).n(r_i+r)>}{<n>} 
$$
which is the average density around an occupied point.
For a fractal medium, where the mass depends on the size as
$$
M(r) \propto r^D
$$ 
$D$ being the fractal (Haussdorf) dimension, the conditional
density behaves as
$$
\Gamma(r) \propto r^{D-3}
$$
This is exactly the statistical analysis used for the interstellar
clouds, since the ISM astronomers have not  adopted from the start 
any large-scale homogeneity assumption (cf Pfenniger \& Combes 1994).

The fact that for a fractal the correlation $\xi(r)$ can be highly
misleading is readily seen since
$$
\xi(r) = \frac{\Gamma(r)}{<n>} -1
$$
and for a fractal structure the average density of the sample $<n>$ is a 
decreasing function of the sample length scale. In the general use of
$\xi(r)$,  $<n>$ is taken for a constant, and we can see that
$$
D = 3 - \gamma \quad .
$$
If for very small scales,
both $\xi(r)$ and $\Gamma(r)$ have the same power-law behaviour, with the 
same slope $-\gamma$, then the slope appears to steepen for $\xi(r)$
when approaching the  length $r_0$. This explains why
with a correct statistical 
analysis (Di Nella et al 1996, Sylos Labini \& Amendola 1996, 
Sylos Labini et al 1996), the actual $\gamma \approx 1-1.5$ is smaller 
than that obtained using $\xi(r)$. This also explains why the amplitude of
$\xi(r)$ and $r_0$ increases with the sample size, and for clusters as well. 

In the following, we adopt the framework of the fractal medium that 
we used for the ISM (dVSC), and will not 
consider any longer $\xi(r)$.

\section{Equations in the comoving frame}

Let us consider the universe in expansion with the characteristic
scale factor $a(t)$. For the sake of simplicity, we modelise the galaxies
by points of equal masses $m$, although they have a mass spectrum
(it may be responsible for a multi-fractal structure, see Sylos Labini
\& Pietronero 1996). 

The present analysis  can be generalised to galaxies of different masses  
following the lines of sec. IX \cite{prd}. We expect to
come to this point in future work.

If the physical coordinates of the particles are  $ {\vec r} $, we can 
introduce the comoving coordinates $ {\vec x} $ such that 
$$
 {\vec r} =  a(t) \; {\vec x} 
$$
The Lagrangian for a system of $ N $ particles interacting only
by their self-gravity can be written as
\begin{equation}\label{lagra}
L_N = \sum_{i=1}^N \left[ \frac{m}2  a(t)^2 \;
 {\dot  {\vec x}}_i^2 - \frac{m}{
a(t)} \; \phi( {\vec x}_i(t)) \right] \; ,
\end{equation}
where  $ \phi( {\vec x}) $ is the gravitational potential in the
comoving frame, determined by the Poisson equation
\begin{equation}\label{poi}
\nabla^2 \phi( {\vec x})= 4\pi G \; \rho( {\vec x}, t) \; ,
\end{equation}
and $  \rho( {\vec x}, t) $ is the mass density. 
For our system of point particles,
\begin{equation}\label{rho}
 \rho( {\vec x}, t) = m \, \sum_{i=1}^N \delta( {\vec x}-  {\vec x}_i(t))
\end{equation}
and therefore the solution of the Poisson equation takes the form
\begin{equation}\label{fi}
 \phi( {\vec x}) = - G m  \sum_{i=1}^N { 1 \over { |  {\vec x}-  {\vec
 x}_i(t)| }}\; .
\end{equation}

The canonical momenta and Hamiltonian of the system are
$$
{\vec p_i} = m\;   a(t)^2  \; {\dot  {\vec x}}_i 
$$
$$
H_N =  \sum_{i=1}^N \left[{ 1 \over { 2 m a(t)^2}}\;{\vec p_i\,}^2 + \frac{m}{
a(t)} \; \phi( {\vec x}_i(t)) \right] 
$$
\begin{equation}\label{hamC}
= { 1 \over { 2 m  a(t)^2}}\; \sum_{i=1}^N {\vec p_i\, }^2 -  {{G \, m^2}
\over {a(t)}} \sum_{1\leq l < j\leq N} {1 \over { |{\vec x}_l - {\vec x}_j|}} 
\end{equation}
We see that the $N$-particle Hamiltonian in cosmological spacetime
eq.(\ref{hamC}) can be obtained from the Minkowski Hamiltonian
[$ a(t)= 1 $] by making the replacements 
\begin{equation}\label{cambios}
m \to m \, a(t)^2 \quad ,  \quad G \to G \,  a(t)^{-5} \; . 
\end{equation}

As a first approximation, we shall assume in the following that   
the characteristic time of the particle motions under the
gravitational self-interaction are shorter than the
time variation of $a(t)$. We can then consider that this
system of self-gravitating particles is at any time 
in approximate thermal equilibrium.
This hypothesis is true of course for structures that have already 
decoupled from the expansion, and are truly self-gravitating and virialised.
 It could be also valid for the whole non-linear regime of
the gravitational collapse. 
As for the linear regime, we know already that the primordial fluctuations
are not forgotten in the large-scale structures, and therefore the
resulting correlations will depend on initial conditions, and not be entirely
determined by self-gravity.

 The above assumption introduces a natural upper limit in the 
scales concerned by the theory developped below. Similarly to the
case of the interstellar medium, the fractal structure considered is
bounded by a short distance cut-off and by a large-scale limit as well
(dVSC).

 The short distance cut-off corresponds to the appearence of other
physics at  short scale, essentially dissipative, which we do not need
to introduce. In addition, the short distance cut-off
avoids the gravo-thermal catastrophe. For the ISM, 
the cut-off was naturally the size of the smaller fragments,
of the order of the Jeans length. Here the cut-off
corresponds also to the size of the `particles' considered, i.e. the
galaxy size, below which another physics steps in, related to
stellar formation and radiation. 

The fact that in the catalogs, we are observing in projection large-scale
structures at different epochs, with different values of the scale
factor $a(t)$, could slightly modify the fractal dimension. Even though fractal
structures are self-similar, and scale-independent, the largest scales
are systematically observed at a younger epoch where the contrast has not
grown up as high as today. This evolution effect however should be significant
only at high redshift ($>1$), and the present catalogs are not yet
statistically robust so far back in time (the average redshift of optical
catalogs is about 0.1).

\section{Application of renormalization group theory}

As in all scale-independent problems, where the fluctuations cannot
be represented by analytical functions, the renormalization group
theory developped in the 1970's for the study of critical phenomena, appears
here perfectly adapted (e.g. \cite{kgw}). We can consider the fractal
structure of the Universe as the critical state of a system, 
where fluctuations develop at any scale, with a very
large correlation length (asymptotically infinite). The fluctuations that are 
distributed as a fractal of dimension $D$ are the large-scale structures of 
the universe (cf. \cite{toki}).

We generalise now the   study of an $N$-body system only interacting through 
their own self-gravity to the case where the bodies are on 
a cosmological background.

Another  approach has been proposed for galaxy correlations
 \cite{pm}, but it yields different critical exponents.

Let us apply the theory to the system of galaxy points, already
defined in the previous section. Since they are considered 
in approximate thermal equilibrium, we will
use the grand canonical ensemble, that also
allows a variable number of particles.
The grand partition function of the system can be written as in eq.{\ref{gfp})
where now $ H_N $ is given by eq.(\ref{hamC}).

The functional representation for the  grand partition function 
can be easily generalized for an arbitrary scale
factor $a(t)$. After the changes specified above in
eq.(\ref{cambios}), the local action becomes
\begin{equation}\label{acciF}
S[\phi(.)] \equiv  { {a(t)} \over{T_{eff}}}
\int d^3x \left[ \frac12(\nabla\phi)^2  - \mu^2   a(t)^2 \;  e^{\phi({\vec
x})}\right] 
\end{equation}
Notice that all quantities depend on time
through the scale factor $ a(t) $ only. There is no integration over $ t $. 

The mass parameter $ \mu $  in the $\phi$-theory gets effectively
multiplied by the scale factor  $ a(t) $.  Since the Jeans length $ d_J
\simeq \mu^{-1} $ according to eq.(\ref{lonJeans}), 
in comoving coordinates $ d_J $ effectively becomes 
$$ 
d_J  =  \sqrt{12\over {\pi}}\; { 1 \over {\mu \, a(t) }} \quad ,
$$ 
as one could have expected.  

On the other hand, the dimensionless coupling constant 
$$
g^2 = \mu \,  T_{eff}
$$
is unchanged by the  replacements of eq.(\ref{cambios}).

Therefore, for any fixed time $ t $  we find the same scaling
behaviour, after making the replacement
$$
\mu \to \mu \; a(t) 
$$
and keeping the coupling $ g $ unchanged. 

Thus, the renormalisation group and finite size scaling analysis of secs.
IVC and IVD apply without essential changes to the galaxy distributions.
Namely, the mass fluctuations $ \Delta M(R) $  inside a volume $ R $,
$$
(\Delta M(R))^2 \equiv  \; <M^2> -<M>^2  \quad ,
$$
will scale as
\begin{equation}
\Delta M(R)  \sim  R^{\frac1{\nu}}\; .
\end{equation}

The scaling exponent $ \nu $ can then be identified as before with the inverse
Haussdorf (fractal) dimension $D$ of the system
$$
D = {1\over{\nu}} \; .
$$

As usual in the theory of critical phenomena, there are only two
independent critical exponents. All exponents can be expressed in
terms of two of them: for instance the fractal dimension $ D = 1/ \nu $,
and the independent exponent $\eta$, 
which usually governs the spin-spin correlation
functions. The exponent  $\eta$ appears here in the $\phi$-field 
correlator (dVSC), describing the gravitational potential, that scales as
$$
<\phi({\vec r})> \; \sim r^{-\frac12(1+\eta)}
$$
The values of the critical exponents depend on the fixed point that governs 
the long range behaviour of the system. 

The value of the dimensionless coupling constant $g^2 = \mu T_{eff}$
should decide whether the fixed point chosen by the system is the
mean field (weak coupling) or the Ising one (strong coupling). 
At the tree level, we estimate  $g \approx \frac{5}{\sqrt{N}}$, where 
$N$ is the number of points in a Jeans volume $d_J^3$. The coupling
constant appears then of the order of $1$, and we cannot settle this question
without effective computations of the renormalisation group equations.
At this point, the predicted fractal dimension $D$ should be between
$1.585$ and $2$.

\subsection{Three point and higher correlations}

Our approach allows to compute higher order correlators without any extra 
assumption.

The two and three point densities,
\begin{eqnarray}
D({\vec r}_1,{\vec r}_2) &\equiv& <n({\vec r}_1)\, n({\vec r}_2)> \cr \cr
D({\vec r}_1,{\vec r}_2,{\vec r}_3) &\equiv& <n({\vec r}_1)\, n({\vec r}_2) \,
n({\vec r}_3)> \; ,
\end{eqnarray}
can be expressed as follows in terms of the correlation functions:
\begin{eqnarray}\label{distr3}
D({\vec r}_1,{\vec r}_2) & = & n_1 \;  n_2 +  C_{12}\cr \cr
D({\vec r}_1,{\vec r}_2,{\vec r}_3) &=& n_1 \;  n_2  \;  n_3 + 
 n_1 \; C_{23} +  n_3 \; C_{12} +  n_2 \; C_{13} + C_{123}\; .
\end{eqnarray}
Here,
$$
n_i \equiv < n({\vec r}_i)> \quad , \; i=1,2,3 \; ,
$$
and $ C_{ij} $ and $ C_{ijk} $ are the two and three point 
correlation functions, respectively,
$$
 C_{ij} \equiv C({\vec r}_i,{\vec r}_j)
$$
$$
 C_{ijk} \equiv C({\vec r}_i,{\vec r}_j,{\vec r}_k)
$$

The behaviour of $ n_i , \;  C_{ij} $ and $ C_{ijk} $ in the scaling regime
follow from the renormalisation group equations at criticality
(de Vega, S\'anchez \& Combes, in preparation). If we
 do not impose homogeneity at all scales, we find,
\begin{eqnarray}\label{rg123}
< n({\vec r})> \simeq A\; r^{D-3} \; , \cr \cr
C({\vec r}_1,{\vec r}_2) \buildrel{r_1 >> r_2 }\over \simeq B\; r_1^{2(D-3)}
\; , \cr \cr
C({\vec r}_1,{\vec r}_2,{\vec r}_3) \buildrel{r_1 >> r_2, r_3 }\over \simeq 
C \;  r_1^{3(D-3)}\; ,
\end{eqnarray}
where $ A, \; B $ and $ C $ are constants and $ D = 1/\nu $.

We can now derive the three point density behaviour when one point, say $ 
{\vec r}_1 $, is far away from the other two. We find from eqs.(\ref{distr3})
and (\ref{rg123}),
\begin{eqnarray}\label{D123}
D({\vec r}_1,{\vec r}_2) &\buildrel{r_1 >> r_2 }\over\simeq &
 A \;  r_1^{D-3} \; n_2 \; + B\; r_1^{2(D-3)} \; , \cr \cr
D({\vec r}_1,{\vec r}_2,{\vec r}_3) 
& \buildrel{r_1 >> r_2, r_3 }\over\simeq & A \;  r_1^{D-3} \; 
\left( n_2 \; n_3 + C_{23} \right) \cr \cr
&+& B\; r_1^{2(D-3)} \; ( n_2 + n_3 ) + C \;  r_1^{3(D-3)}
\end{eqnarray}
Notice that this expression is dominated by the first term since $ D- 3 < 0 $.

Higher point distributions can be treated analogously in our approach. 
We find that the dominant behaviour in the $N$-points density is 
\begin{equation}\label{Ncorre}
C({\vec r}_1,{\vec r}_2,\ldots,{\vec r}_N) 
 \buildrel{r_1 >> r_i, \; 2\leq i \leq N  }\over\sim r_1^{N(D-3)} 
\end{equation}
Notice that when homogeneity is assumed to hold over all scales, the critical 
behaviour of the $N$-point correlation function involves a factor
$ r_1^{(N-1)(D-3)} $, \cite{itdr}.

Eqs.(\ref{D123}-\ref{Ncorre}) are qualitatively similar, although
 not identical, to the behaviour inferred assuming the factorized hierarchical 
Ansatz (fhA), (Balian \& Schaeffer 1989). That is,
\begin{eqnarray}\label{fhA}
D({\vec r}_1,{\vec r}_2)^{fhA} &=& {\bar n}^2 \left(1 + 
b\;  r_{12}^{D-3} \right) \buildrel{r_1 >> r_2 }\over\simeq 
 {\bar n}^2 \left(1 + r_{1}^{D-3} \right)\; , \\ \cr
D({\vec r}_1,{\vec r}_2,{\vec r}_3) ^{fhA} &=&  {\bar n}^3 \left\{
1 +  b\;\left(   r_{12}^{D-3} +  r_{13}^{D-3} + r_{23}^{D-3} \right) \right.
\cr \cr
&+&  \left. Q_3 \left[ r_{12}^{D-3}\;   r_{13}^{D-3} + 
r_{12}^{D-3}\;r_{23}^{D-3} + r_{13}^{D-3}\;   r_{23}^{D-3} \right] \right\}
\cr \cr& \buildrel{r_1 >> r_2, r_3 }\over\simeq &
{\bar n}^3 \left[ 1 +  b\; r_{23}^{D-3} + 2 \,  r_{1}^{D-3} \,
( b +  Q_3\;r_{23}^{D-3}) +  Q_3 \; r_1^{2(D-3)} \right]
\; . \nonumber
\end{eqnarray}
where  $ r_{12} \equiv |{\vec r}_1 - {\vec r}_2 | $ and so on.
$ b $ and $ Q_3 $ are  constants. Notice that in the factorized hierarchical 
Ansatz, the fractal dimension $ D $ is not predicted but it is a free 
parameter.

We see that the dominant behaviours in eqs.(\ref{D123}) and  (\ref{fhA})
are similar in case  the  scaling exponents $ D - 3 $ are the same.

\section{Discussion}

In previous sections we ignored gravitational forces external to the
gas like stars etc. Adding a  fixed external mass density $
\rho_{ext}({\vec r}) $ amounts to introduce an external source
$$
J({\vec r}) =  - T_{eff}\; \rho_{ext}({\vec r})\; ,
$$
in eq.(\ref{zfiJ}). Such term will obviously affect correlation
functions, the  mass density, etc. except when we look at the scaling
behaviour which is governed by the critical point. 
That is, the values we find for the scaling exponents $ d_H $ and $ q
$ are {\bf stable} under external perturbations.

\bigskip

We considered all atoms with the same mass in the gravitational gas.
It is easy to generalize the transformation into the $\phi$-field
presented in section II for a mixture of several kinds of atoms. Let
us consider $ n $ species of atoms with 
masses $ m_a, \; 1 \leq a \leq n $. Repeating the steps from
eq.(\ref{gfp}) to (\ref{acci})
yields again a field theory with a single scalar field  but the
action now takes the form
\begin{equation}\label{gasM}
S[\phi(.)] \equiv  {1\over{T_{eff}}}\;
\int d^3x \left[ \frac12(\nabla\phi)^2 \; - \sum_{a=1}^n \; \mu_a^2 \;
e^{{{m_a}\over m}\, \phi({\vec x})}\right] \; ,
\end{equation}
where
$$
\mu_a^2 = \sqrt{2\over {\pi}}\; z_a \; G \, m_a^{3/2} \, m^2 \, \sqrt{T}
\; ,
$$
and $ m $ is just a reference mass. 

Correlation functions, mass densities and other observables will 
obviously depend on the number of species, their masses and fugacities
but it is easy to see that the fixed points and scaling exponents are
exactly the {\bf same} as for the $\phi$-field theory (\ref{zfi})-(\ref{muyT}).

\bigskip

We want to notice that there is an important difference between the
behaviour of the gravitational gas and the spin models (and all other
statistical models in the same universality class). For the
gravitational gas we find scaling behaviour for a {\bf full range} of
temperatures and couplings. For spin models scaling only appears 
at the critical value of the temperature. At the critical temperature
 the correlation length $ \xi $ is infinite and the theory is
massless. 
For temperatures near the critical one,  i. e. in the critical
domain,  $ \xi $ is finite (although very
large compared with the lattice spacing) and the correlation functions
decrease as $ \sim e^{ - r/\xi} $ for large distances $ r $.  
Fluctuations of the relevant operators support perturbations which can
be interpreted as massive excitations. Such
(massive) behaviour does not appear for the gravitational gas. The ISM
correlators scale exhibiting power-law behaviour. This feature is
connected with the scale invariant character of the Newtonian force
and its infinite range.

\bigskip

The hypothesis of strict thermal equilibrium does not apply to the ISM as 
a whole where temperatures range from $ 5 $ to $ 50 $ K and even $ 1000 $ K. 
However, since the scaling behaviour is independent of the temperature,
it applies to {\bf each} region of the ISM in thermal equilibrium.
Therefore, our theory applies provided thermal equilibrium holds
in regions or clouds. 

 We have developped here the theory of a gravitationally interacting
ensemble of bodies at a  fixed temperature.  In a real situation like the ISM, 
 gravitational perturbations  from external masses,
as well as other perturbations are present.
We have shown that the scaling solution is stable
with respect to the gravitational perturbations. It is well known that 
solutions based on a fixed point are generally quite robust.

Our theory  especially applies  to the interstellar medium far from 
star forming regions, which can be locally far from thermal equilibrium,
and where ionised gas at 10$^4$K together with coronal gas at 10$^6$K 
can coexist with the cold interstellar medium. In the outer parts of
galaxies, devoid of star formation, the ideal isothermal conditions 
are met \cite{pcm}. Inside the Galaxy, large regions satisfy also
the near isothermal criterium, and these are precisely the regions
where scaling 
laws are the best verified. Globally over the Galaxy, the fraction
of the gas in the hot ionised phase represents a negligible mass, 
a few percents, although occupying a significant volume. Hence, this
hot ionised gas is a perturbation which may not  change the fixed point
behaviour of the thermal  self-gravitating gas.

\bigskip

In ref.\cite{pm} a connection between a gravitational gas of galaxies 
in an expanding universe and the
Ising model is conjectured. However, the unproven identification made  
in ref.\cite{pm} of the  mass density contrast with the Ising spin leads to
scaling exponents  different from ours.

\bigskip

\bigskip

Concerning the galaxy distributions, 
different scaling behaviours show up depending whether
the system is homogeneous or not at large distances. The
homogeneity property brings  extra information which is not contained
in the fundamental gravitational interaction. In condensed matter
systems, such homogeneity hypothesis is easily verified by experiments.
The homogeneity of the universe at large scales is a much more
controversial issue \cite{peeb,pietro}.

Under the homogeneity hypothesis we find for the galaxy-galaxy
correlator defined as in \cite{peeb},
$$
\xi(r) \equiv {{<\rho({\vec r_0})\rho({\vec r_0} + {\vec r}) >}\over 
{<\rho>^2}}-1 \sim r^{-\gamma}
$$
where $ \gamma = 6 - 2 d_H $.  This gives $ \gamma_{mean\, field} = 2
$ and  $ \gamma_{Ising} = 2.830\ldots $. Such numbers should be
compared with the customary  value  $ \gamma = 1.8 $ obtained from the
observations \cite{peeb}.

In the inhomogeneous regime we find for the galaxy-galaxy correlations
considered in ref.\cite{pietro},
$$
D(r) = <\rho({\vec r_0})\rho({\vec r_0} + {\vec r}) > \; \sim\;  r^{-\Gamma}
$$
where $ \Gamma = 3 - d_H $. This gives $ \Gamma_{mean\, field} = 1 $
and $ \Gamma_{Ising} = 1.415\ldots $.  Such numbers should be
compared with the observed value for $ \Gamma \simeq 1 $ \cite{pietro}
(obtained for $ r > 30 \, h^{-1}\,$ Mpc). 

In conclusion, our renormalization group results seem to fit better with the
analysis of the observations made in ref.\cite{pietro} than with the
standard lore\cite{peeb}. However, our predictions do not disagree in
a dramatic way with the standard analysis of the observations.
In both cases, mean field exponents give the better agreement with
observations.

Nuevas conclusiones:

 The statistical analysis of the most recent galaxy catalogs,
without the assumption of homogeneity at a scale smaller than the
catalog depth, has determined that the universe has a fractal structure
at least up to $\approx 100 h^{-1}$ Mpc (Sylos Labini et al 1996). 
The analysis in terms of conditional density has revealed that
the fractal dimension is between $D$ = 1.5 and 2
(Di Nella et al 1996, Sylos Labini \& Amendola 1996). We apply a
theory that we have developped to explain the fractal structure
of the interstellar medium (dVSC), which has the same dimension $D$.
The physics is based on the self-gravitating interaction of an
ensemble of particles, over scales limited both at short and
large distances. The short-distance cut-off is brought by other
physical processes including dissipation.
The long-range limit is fixed by the expansion time-scale.
In-between, the system is assumed in approximate thermal equilibrium.
  The dynamical range of scales involved in this thermodynamic
quasi-equilibrium is at present limited to 3-4 orders of magnitude,
but will increase with time. 

 The critical exponents found
in the theory do not depend on the conditions at the cut-off, which
determine only the amplitudes. The theory is
based on the statistical study of the gravitational field: it is shown
that the partition function of the N-body ensemble is equivalent to 
the partition function of a single scalar field, with a local action.
This allows to use field theory methods and the renormalisation group
to find the scaling behaviour. We find scaling behaviour for a {\bf
full range} of temperatures and couplings.
The theory then predicts for the system a
fractal dimension $D= 1.585$ for the Ising fixed point,
or $D=2$ in the case of the mean-field fixed point. Both are compatible
with the available observations. The $N$-points density correlators
are predicted to scale with exponent $(N-1)(D-3)$ when $ r_1 >> r_i, \;
2\leq i \leq N $. That is, $ -(N-1) $ for the mean field, or $ -1.415\,(N-1) $
for the Ising point.

We predict in addition a critical exponent
$ - \frac12 (1 + \eta ) $ for the gravitational potential: that is, $ -0.500
$ for mean field or $ - 0.519 $ for the Ising fixed point.


\end{document}